\documentclass[%
 preprint,
superscriptaddress,
 amsmath,amssymb,
 aps,
prb,
]{revtex4-1}

\usepackage[utf8] {inputenc}
\usepackage[english] {babel}
\usepackage{graphicx}
\usepackage{dcolumn}
\usepackage{bm}
\usepackage{mathtools}
\usepackage{float}
\usepackage{multirow}
\usepackage{amsmath}

\usepackage{lmodern}
\usepackage{textcomp}
\usepackage{t1enc}

\usepackage{graphicx}
\usepackage{ulem}
\usepackage{setspace}

\usepackage[toc, page]{appendix}
\usepackage{standalone}

\usepackage{verbatim}

\usepackage{natbib}

\newlength{\depthofsumsign}
\newlength{\totalheightofsumsign}
\newlength{\heightanddepthofargument}

\newcommand{\nsum}[1][1.4]{
    \mathop{%
        \raisebox
            {-#1\depthofsumsign+1\depthofsumsign}
            {\scalebox
                {#1}
                {$\displaystyle\sum$}%
            }
    }
}

\makeatletter
\newcommand*{\DivideLengths}[2]{%
  \strip@pt\dimexpr\number\numexpr\number\dimexpr#1\relax*65536/\number\dimexpr#2\relax\relax sp\relax
}
\makeatother

\begin{document}


\title{Fluorescence spectrum and charge state control of divacancy qubits via illumination at elevated temperatures in 4H silicon carbide}

\author{A. Cs\'or\'e}
\affiliation{%
 Department of Atomic Physics, Budapest University of Technology and Economics, Budafoki út 8., H-1111, Budapest, Hungary}

\author{I. G. Ivanov}
\affiliation{%
Department of Physics, Chemistry and Biology, Link\"oping University, SE-58183 Link\"oping, Sweden}

\author{N. T. Son}
\affiliation{%
Department of Physics, Chemistry and Biology, Link\"oping University, SE-58183 Link\"oping, Sweden}

\author{A. Gali}
 \affiliation{%
 Department of Atomic Physics, Budapest University of Technology and Economics, Budafoki út 8., H-1111, Budapest, Hungary}
\affiliation{%
 Wigner Research Centre for Physics, PO. Box 49, Budapest H-1525, Hungary}

\date{\today}

\begin{abstract}

Divacancy in its neutral charge state (V$_\text{C}$V$_\text{Si}^0$) in 4H silicon carbide (SiC) is a leading quantum bit (qubit) contender. Owing to the lattice structure of 4H SiC, four different V$_\text{C}$V$_\text{Si}$ configurations can be formed. The ground and the optically accessible excited states of V$_\text{C}$V$_\text{Si}^0$ configurations exhibit a high-spin state, and the corresponding optical transition energies are around $\approx 1.1$~eV falling in the near-infrared wavelength region. Recently, photoluminescence (PL) quenching has been experimentally observed for all V$_\text{C}$V$_\text{Si}$ configurations in 4H SiC at cryogenic temperatures. It has been shown that V$_\text{C}$V$_\text{Si}^0$ is converted to V$_\text{C}$V$_\text{Si}^-$ and it remains in this shelving dark state at cryogenic temperatures until photoexcitation with the threshold energies or above is applied to convert V$_\text{C}$V$_\text{Si}^-$ back to V$_\text{C}$V$_\text{Si}^0$. In this study, we demonstrate both in experiments and theory that the threshold energy for reionization is temperature dependent. We carry out density functional theory (DFT) calculations in order to investigate the temperature dependent reionization spectrum, i.e., the spectrum of the V$_\text{C}$V$_\text{Si}^- \rightarrow$ V$_\text{C}$V$_\text{Si}^0$ process. We find that simultaneous optical reionization and qubit manipulation can be carried out at room temperature with photoexcitation at the typical excitation wavelength used for readout of the divacancy qubits in 4H SiC, in agreement with our experimental data. We also provide the analysis of the PL spectrum of V$_\text{C}$V$_\text{Si}^0$, characteristic for each V$_\text{C}$V$_\text{Si}^0$ configuration in 4H SiC, using the Huang-Rhys theory, and find that one configuration in 4H SiC stands out in terms of the strength of coherent emission among the four configurations. 

\end{abstract}

\pacs{Valid PACS appear here}
\maketitle

\section{\label{intro}Introduction}

Spin active point defects in wide band gap semiconductors have proven to be promising candidates for numerous applications in the rapidly emerging field of quantum technology.  Leading contender of this family is the negatively charged nitrogen-vacancy defect, i.e., the NV center in diamond with exhibiting many desirable magneto-optical properties~\cite{DohertyPhysRep2013, Gali2019}. Nevertheless, further alternative point defect color centers have been searched for in the past decade embedded in technologically more mature materials than diamond.

A suitable host candidate to this end is silicon carbide (SiC) introducing different crystal modifications called polytypes. The technologically most relevant polytype is the hexagonal 4H SiC which hosts different types of quantum bits (qubits)~\cite{KoehlNature2011, CastelettoNatMater2013, ChristleNatMater2015, Widmann2015NatMat, Wolfowicz2020} being pivotal in quantum information processing~\cite{LohrmannNatComm2015, RadulaskiNanoLett2017, SpindlbergerPRA2019, Nagy2019, Babin2022}, ultrasensitive nanosensors in magnetometry~\cite{SangYunPRB2015, KrausSciRep2014, Simin2016PRX, SiminPRApplied2015, NiethammerPRApplied2016, CochraneSciRep2016} or thermometry~\cite{KrausSciRep2014, AnisimovSciRep2016}.

\begin{figure*} [t]
\includegraphics[width=\textwidth]{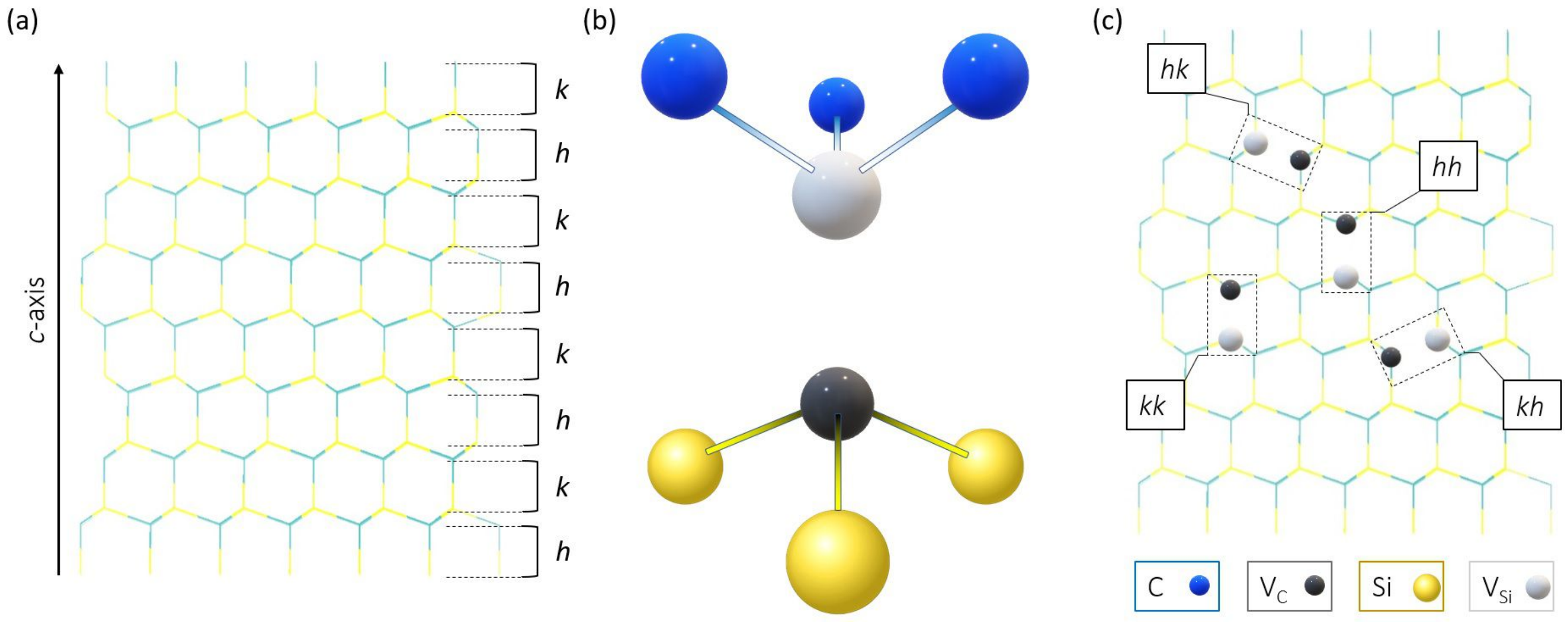}
\caption{(a) Lattice structure of 4H SiC. The [0001] crystal axis ($c$-axis) and the $h$/$k$ bilayers are depicted. (b) Microscopic structure of the divacancy defect. (c) The four possible V$_\text{C}$V$_\text{Si}$ configurations embedded into 4H SiC. Color code of the atoms and the vacant sites is indicated.}
\vspace{0 pt}
\label{fig:struct}
\end{figure*}

Divacancy is one of the most successful qubits in 4H SiC~\cite{Gali2011, KoehlNature2011} which consists of neighbor carbon and silicon vacancies, V$_\text{C}$V$_\text{Si}$. 4H SiC consists of Si-C bilayers exhibiting hexagonal ($h$) or quasicubic ($k$) sites. Consequently, V$_\text{C}$V$_\text{Si}$ possesses four different defect configurations in 4H SiC because of the two possible defect sites for both V$_\text{C}$ and V$_\text{Si}$. Henceforward, we denote the different configurations with the lattice sites' labels of the C and Si vacancies, respectively, such as V$_\text{C}$V$_\text{Si}$($hh$), V$_\text{C}$V$_\text{Si}$($kk$), V$_\text{C}$V$_\text{Si}$($hk$) and V$_\text{C}$V$_\text{Si}$($kh$). The $hh$ and $kk$ configurations are called axial or on-axis configurations exhibiting C$_\text{3v}$ symmetry, while the $hk$ and $kh$ configurations are basal or off-axis configurations with C$_\text{1h}$ symmetry. Lattice structure of 4H SiC and the V$_\text{C}$V$_\text{Si}$ defect configurations are shown in Fig.~\ref{fig:struct}.

The divacancy in 4H SiC exhibits several desirable magneto-optical properties making it a promising qubit. In particular, electronic structure of the neutral V$_\text{C}$V$_\text{Si}$ center (V$_\text{C}$V$_\text{Si}^0$) introduces high-spin ($S=1$) ground and optically accessible excited states~\cite{Magnusson2005, BaranovJETP2005, SonPRL2006, KoehlNature2011}. The corresponding no-phonon optical transitions, i.e., the zero-phonon lines (ZPLs) are around 1100~nm falling into the near-infrared (NIR) region where optical fibers and organic tissues have relatively low absorption window. This property makes the divacancy a promising candidate as a single photon emitter with quantum memory for quantum communication and as an ultrasensitive biomarker~\cite{SomogyiIOP2014, SomogyiNanoscale2012, BekeJPCL2020}. Indeed, high-fidelity infrared spin-photon interface~\cite{ChristlePRX2017} and entaglement and control of single nuclear spins~\cite{BourassaNatMat2020} have been demonstrated with divacancies in 4H SiC. The ZPLs are now labeled~\cite{KoehlNature2011} and identified with the corresponding configurations~\cite{Falk2014} as PL1 - V$_\text{C}$V$_\text{Si}$($hh$), PL2 - V$_\text{C}$V$_\text{Si}$($kk$), PL3 - V$_\text{C}$V$_\text{Si}$($hk$) and PL4 - V$_\text{C}$V$_\text{Si}$($kh$) that were previously labeled as UD2 centers in 4H SiC in the literature~\cite{Magnusson2005}.  

The charge state control of defect qubits is indispensable for robust quantum information processing~\cite{Wolfowicz2021}. Photoexcitation of target point defect qubits may lead to their unwanted charge switching with losing the qubit state. According to previous \textit{ab initio} studies~\cite{GaliMSF2006, GaliJMR2012, GordonPRB2015, MagnussonPRB2018}, all the divacancy configurations could appear in the positive, neutral, negative and double negative charge states depending on the position of the Fermi-level in 4H SiC which may be actively controlled by doping and electrical switching in a 4H SiC diode~\cite{Anderson2019}. Nevertheless, photoexcitation of the neutral divacancies may lead to the ionization of the defect. Indeed, photoluminescence (PL) quenching of divacancy defects in 4H SiC has been recently observed, i.e., the PL signals of V$_\text{C}$V$_\text{Si}^0$ defects do not appear upon certain excitation energies~\cite{ZwierSciRep2015, WolfowiczNatComm2017, GolterSciRep2017, MagnussonPRB2018}. On the other hand, it has been demonstrated that applying a second laser (repump laser) with a higher photon energy completely recovers the PL intensity of the V$_\text{C}$V$_\text{Si}^0$ centers~\cite{ZwierSciRep2015, WolfowiczNatComm2017, GolterSciRep2017, MagnussonPRB2018}. Here, we note that the PL quenching phenomenon is not unique for V$_\text{C}$V$_\text{Si}^0$ in 4H SiC but was observed for several other defects (e.g., see Ref.~\onlinecite{Gali2019} and references therein for NV center in diamond). The physical mechanism behind the quenching phenomenon may be the charge state switching of the corresponding defect upon illumination as already demonstrated by previous experimental and theoretical works~\cite{ZwierSciRep2015, WolfowiczNatComm2017, GolterSciRep2017, SiyushevPRL2013, MagnussonPRB2018}. In this case, the role of the repump laser is to reionize the defect, i.e., to maintain the neutral charge state. In particular, the 'dark' charge state has been attributed to either the singly positive~\cite{GolterSciRep2017} or the singly negative charge state~\cite{WolfowiczNatComm2017} of divacancies by earlier experimental works. In order to unravel this issue, we carried out a combined experimental and density functional theory (DFT) study~\cite{MagnussonPRB2018} and found that the singly negative charge state [V$_\text{C}$V$_\text{Si}^-$] is the dark state because the calculated and observed photoexcitation threshold energies at cryogenic temperatures agreed well for this charge state for all the divacancy configurations in 4H SiC.

In this work, we investigate the temperature dependent charge state control of the V$_\text{C}$V$_\text{Si}^0$ defect configurations in 4H SiC. To this end, we calculate the V$_\text{C}$V$_\text{Si}^0$-related PL lineshape and the V$_\text{C}$V$_\text{Si}^-$ $\rightarrow$ V$_\text{C}$V$_\text{Si}^0$ reionization spectra based on the Huang-Rhys (HR) theory~\cite{HuangProcRoySoC1950, AlkauskasNJP2014, GaliNatComm2016}, and monitor the apperance of the divacancy PL spectra as a function of the applied repump laser's wavelength and temperature in high-purity semi-insulating 4H SiC samples. We find an excitation scheme with simultaneous charge state control and qubit manipulation of V$_\text{C}$V$_\text{Si}^0$ by a single color laser beam at elevated temperatures. Our paper is organized as follows: we describe the computational methodology in Sec.~\ref{sec:method}, where we briefly describe the HR theory in Sec.~\ref{subsec:HRmethod}, the temperature dependence of the charge transition levels in Sec.~\ref{subsec:CTLTdep} and the methodology of the \textit{ab initio} DFT calculations in Sec.~\ref{subsec:compmethod}. We report and discuss our results in Sec.~\ref{sec:ResDisc}. In particular, PL spectra at cryogenic temperature of the V$_\text{C}$V$_\text{Si}^0$ defect configurations are discussed in Sec.~\ref{subsec:HRPL}, the simulated V$_\text{C}$V$_\text{Si}^-$ $\rightarrow$ V$_\text{C}$V$_\text{Si}^0$ temperature dependent reionization spectra are reported within various levels of approximation in Secs.~\ref{subsec:HRre}-\ref{subsec:FreeE}. The experimental data are presented and interpreted by our DFT simulations in Sec.~\ref{subsec:Exp}. We conclude our work in Sec.~\ref{sec:Concl}.


\section{\label{sec:method}Methodology}

\subsection{\label{subsec:HRmethod} Huang-Rhys theory}

The PL spectrum [$I^\text{EM}(\hbar\omega)$] within HR theory can be expressed as~\cite{AlkauskasNJP2014}
\begin{equation}
I^{\text{EM}}(\hbar\omega)=C^{\text{EM}}\omega^{3}\mu_{\text{EG}}^{2}\sum_{n,m}c_{m}(T)\Big|\Big\langle\Theta_{\text{ES}}^{m}(Q)\Big|\Theta_{\text{GS}}^{n}(Q)\Big\rangle\Big|^{2}\delta(E_{\text{ES}}^{m}-E_{\text{GS}}^{n}-\hbar\omega)\text{,}
\label{eq:spectPL}
\end{equation}
where the transitions from the corresponding vibronic substates of the excited state (ES), indexed by $m$, to those of the ground state (GS) with the index $n$ are also considered establishing the phonon sideband (PSB) besides the ZPL, i.e., $m=n=0$ in $I^\text{EM}(\hbar\omega)$ yields the intensity of ZPL. Generalized coordinate $Q$ appears in both the ES and GS vibrational function $\Theta$ according to the HR theory. The temperature dependence appears by the factor $c_m(T)$ with temperature $T$ which will be defined later. In Eq.~\ref{eq:spectPL} $\pmb{\mu}_\text{EG}$ is the corresponding matrix element (which is now a three component vector with the length of $\mu_\text{EG}$) of the transition dipole moment operator, $\pmb{\hat{\mu}}_\text{EG} = q\sum_j\mathbf{\hat{r}}_j$ defined as
\begin{equation}
\pmb{\mu}_\text{EG} = \Big\langle \Psi_\text{ES}(\mathbf{r}_i) \Big|\pmb{\hat{\mu}}_\text{EG} \Big| \Psi_\text{GS}(\mathbf{r}_i) \Big\rangle
\label{eq:transdip}
\end{equation}
and $\hbar\omega$ is the photon energy. The prefactor $\omega^3$ in Eq.~\ref{eq:spectPL} consists of the photon density of states (DOS) causing spontaneous emission ($\sim\omega^2$) and the perturbing field of those photons ($\sim\omega$). In contrast, the absorption spectrum is a linear function of $\omega$ since no spontaneous emission is involved. Thus the absorption spectrum can be expressed as
\begin{equation}
I^\text{ABS}(\hbar\omega) = C^\text{ABS} \mu_\text{EG}^2  \omega \nsum[2]_{n,m} c_n(T)\Big| \Big\langle\Theta^{n}_\text{GS}(Q)\Big|\Theta^m_\text{ES}(Q)\Big\rangle \Big|^2\delta(E^m_\text{ES}-E^{n}_\text{GS}-\hbar\omega)\text{,}
\label{eq:spectabs}
\end{equation}
where $C^\text{EM}$ and $C^\text{ABS}$ are the corresponding material and measurement dependent constants of the emission and absorption in Eqs.~\ref{eq:spectPL}~and~\ref{eq:spectabs}, respectively. Since $C^\text{EM/ABS}$ and $\pmb{\hat{\mu}}_\text{EG}$ are constants across the spectrum, we used Eqs.~\ref{eq:linem}~and~\ref{eq:linabs} as the luminescence and absorption lineshapes [$L^\text{EM/ABS}(\hbar\omega)$], respectively, defined as
\begin{equation}
L^\text{EM}(\hbar\omega) =  \omega^3 \nsum[2]_{n,m} c_m(T) \Big| \Big\langle\Theta^{n=0}_\text{GS}(Q)\Big|\Theta^m_\text{ES}(Q)\Big\rangle \Big|^2\delta(E^m_\text{ES}-E^{n}_\text{GS}-\hbar\omega)
\label{eq:linem}
\end{equation}
and
\begin{equation}
L^\text{ABS}(\hbar\omega) =  \omega \nsum[2]_{n,m} c_n(T)  \Big| \Big\langle\Theta^{n}_\text{GS}(Q)\Big|\Theta^m_\text{ES}(Q)\Big\rangle \Big|^2\delta(E^m_\text{ES}-E^{n}_\text{GS}-\hbar\omega) \text{.}
\label{eq:linabs}
\end{equation}

The vibrational overlap integral can be calculated by applying the HR theory~\cite{HuangProcRoySoC1950} as implemented previously~\cite{GaliNatComm2016}. In the HR framework three basic assumptions are used: ($i$) the normal modes and ($ii$) the vibrational frequencies are identical in the initial and final electronic states, and ($iii$) the equilibrium configuration is shifted by $\Delta \mathbf{Y}$ in the final state with respect to the initial nuclear configuration. Generalized nuclear configuration weighted by the nuclear masses ($\mathbf{Y}$) can be defined as $\mathbf{Y} = \mathbf{M}^{\frac{1}{2}}\mathbf{X}$, where $\mathbf{X} = (\mathbf{R}_1,...,\mathbf{R}_N) = (X_1, Y_1, Z_1,..., X_N, Y_N, Z_N)$ is a vector constructed by the nuclear coordinates and
\begin{equation}
\mathbf{M} = \begin{bmatrix}
M_1 & & & & &\\
& M_1 & & & &\\
& & M_1 & & &\\
& & & M_2 & &\\
& & & & \ddots &\\
& & & & & M_N
\end{bmatrix}
\label{eq:masstens}
\end{equation}
is the mass tensor. The partial HR factor ($S_k$) can be calculated as the scalar product of the normal vectors of mode $k$ ($\mathbf{Y}_0^k$) and the displacement vector of $\Delta \mathbf{Y}$ as 
\begin{equation}
S_k = (\Delta \mathbf{Y}^\text{T}\mathbf{Y}_0^k)^2\text{.}
\label{eq:partialS}
\end{equation}
The (total) HR factor ($S$) can be defined as the sum of the partial ones,
\begin{equation}
S = \sum_k^{3N-6}S_k\text{.}
\label{eq:totalS}
\end{equation}
While the partial HR factor indicates the weight of phonon mode $k$ in the corresponding electronic transition, total HR factor gives a measure for the strength of the overall electron-phonon coupling. By using the HR factors, overlapping integral of the initial and final vibrational wavefunctions describing the lineshape of the spectrum can be written as
\begin{equation}
\Big\langle\Theta^m_\text{f}(Q)\Big|\Theta^n_\text{i}(Q)\Big\rangle = \prod_k \langle m_k | n_k \rangle,
\label{eq:overlapHR}
\end{equation}
where $\Theta^m_\text{i}(Q)$ and $\Theta^n_\text{f}(Q)$ stand for the vibrational wavefunctions of the initial and final states, respectively  $m_k$ and $n_k$ stand for the phonon occupation of mode $k$ in the final and initial states with the corresponding $n$ and $m$ quantum numbers, respectively, and
\begin{equation}
 \Big| \langle m_k | n_k \rangle \Big| = e^{-S_k}\frac{S_k^{m-n}}{(m-n)!} \text{.}
\label{eq:modeoverlapHR}
\end{equation}

At $T=0$~K, the lowest energy phonon state is occupied in the electronic excited state in the PL process, i.e., $m=0$. As the temperature raised higher energy $m$ phonon states are occupied. Boltzmann distribution function may be applied at a given temperature for calculating the occupation of $\omega_k$ phonon state which is given by $c_m(T)$ in Eqs.~\ref{eq:spectPL} and \ref{eq:linem}. By defining $\varepsilon$ as the vibronic excited state's ($m$) energy in the electronic excited state,
\begin{equation}
\varepsilon_{m}=E_{\text{ES}}^{m}-E_{\text{ES}}^{0}\text{,}
\end{equation}
$c_m(T)$ can be given as
\begin{equation}
c_{m}=\frac{\exp\Bigl(\frac{-\varepsilon_{m}}{k_{B}T}\Bigr)}{\sum_{m}\exp\Bigl(\frac{-\varepsilon_{m}}{k_{B}T}\Bigr)}\text{.}
\end{equation}
By this way, the temperature dependent PL lineshape may be obtained. In the case of absorption spectrum, $c_n(T)$ can be defined analogously to $c_m(T)$ but it rather represents the occupation of the vibronic excites state in the electronic ground state. We note that all HR spectra presented in this work are normalized.  

Experimental determination of the $S$ factor can be carried out via the so-called Debye-Waller (DW) factor~($w$)~\cite{DebyeAdP1913, WallerZfP1923} which can be directly read out from the corresponding spectra. DW factor represents the ratio of the ZPL and total intensity in the corresponding spectrum and can be expressed using $S$ as $w = \text{exp}(-S)$. In this work, we present normalized HR spectra, i.e., probability density functions (PDFs) of the corresponding electronic transition. By this definition, the area under the curve is one in all cases, and thus the DW factor is the area under the ZPL peak. We note that the temperature dependent width of ZPL emission is not incorporated in this theory and the long wavelength acoustic phonons are not sampled in our calculations due to the finite size of the supercell.

\subsection{\label{subsec:CTLTdep} Charge transition levels}

Formation energy as defined in Ref.~\onlinecite{CsorePRB2021} can be expressed for the V$_\text{C}$V$_\text{Si}$ defect as
\begin{equation}
E_{\text{form}}^q = E_{\text{tot}}^q - \frac{n_{\text{Si}} + n_{\text{C}}}{2}\mu_{\text{SiC}} + qE_\text{F} + \Delta V(q),
\label{eq:formE}
\end{equation}
where $E_{\text{tot}}^q$ is the total energy of the defective system in the $q$ charge state, $n_\text{Si/C}$ is the number of the Si/C atoms in the supercell, $\mu_\text{SiC}$ is the chemical potential of Si-C unit in the perfect 4H SiC crystal, $E_\text{F}$ represents the Fermi-level and $\Delta V(q)$ stands for the charge correction term. To determine $\Delta V(q)$, we use the Freysoldt charge correction scheme~\cite{FreysoldtPRL2009}. For calculating the $(-/0)$ charge transition level referenced to the conduction band minimum (CBM) one may use the following formula
\begin{equation}
E^\text{CBM}_{q+1/q} = E_\text{CBM} - E_{q+1/q}\text{,}
\label{eq:CTLCBM}
\end{equation}
where $E_\text{CBM}$ is the CBM energy, and $E_{q+1/q}$ is the charge transition level calculated as
\begin{equation}
E_{q+1/q} = E_{\text{tot}}^q - E_{\text{tot}}^{q+1} + \Delta V(q) - \Delta V(q+1)\text{.}
\label{eq:CTLs}
\end{equation}

By increasing the temperature, the charge transition levels may shift due to the thermal evolution of the band edges and the thermal shift of the vibrational free energy. 

Effect of the thermal shift of the CBM manifests in Eq.~\ref{eq:CTLCBM}, where $E_\text{CBM}$ is in fact temperature dependent. Since $E_\text{CBM}$ decreases upon increasing the temperature~\cite{CannucciaPRMat2020}, the $(-/0)$ charge transition level referenced to the CBM and hence the reionization threshold energy will also decrease upon introducing this correction as $E^{\text{CBM}^*}_{-/0} = E^\text{CBM}_{-/0} - \Delta E^\text{CBM}(T)$.

On the other hand, thermal shift of the charge transition levels originating from the vibrational entropy of the defect is governed by the temperature dependent vibrational free energy in the corresponding charge state ($F^q$) formulated as~\cite{WickamaranteAPL2018}
\begin{equation}
F^q(T) = \sum_i\Big\{ \frac{1}{2} \hbar\omega_i + k_\text{B}T\text{ln}\Big[1-\text{exp}\Big(-\frac{\hbar\omega_i}{k_\text{B}T}\Big)\Big]\Big\},
\label{eq:freeE}
\end{equation}
where $\hbar$ and $k_\text{B}$ are the reduced Planck constant and Boltzmann constant, respectively, and $\omega_i$ is the frequency of the $i$th phonon mode. Since free energy contributes to the total energy of the system, the $E_{-/0}$ charge transition level shifts by the free energy difference of the negative and neutral charge states [$\Delta F(T)$]. The free energy difference is defined as $\Delta F(T) = F^-(T) - F^0(T)$, so the free energy-shifted charge transition level can be written as $E^{\text{CBM}^{**}}_{-/0} = E^\text{CBM}_{-/0} + \Delta F(T)$. In order to calculate $\Delta F(T)$, we considered all the quasilocal phonon modes associated with the motion of the atoms within the second neighbor shells around the V$_\text{C}$V$_\text{Si}$ defect both in the negative and neutral charge states.  

Combining the two correction terms, the reionization energy threshold energy is defined as
\begin{equation}
E^{\text{CBM}}_{-/0}(\text{corr}) = E^{\text{CBM}}_{-/0} -  \Delta E^\text{CBM}(T) + \Delta F(T)\text{.}
\label{eq:Ereicorr}
\end{equation}

\subsection{\label{subsec:compmethod} Computational methodology}

\begin{figure} [b]
\includegraphics[width=0.5\textwidth]{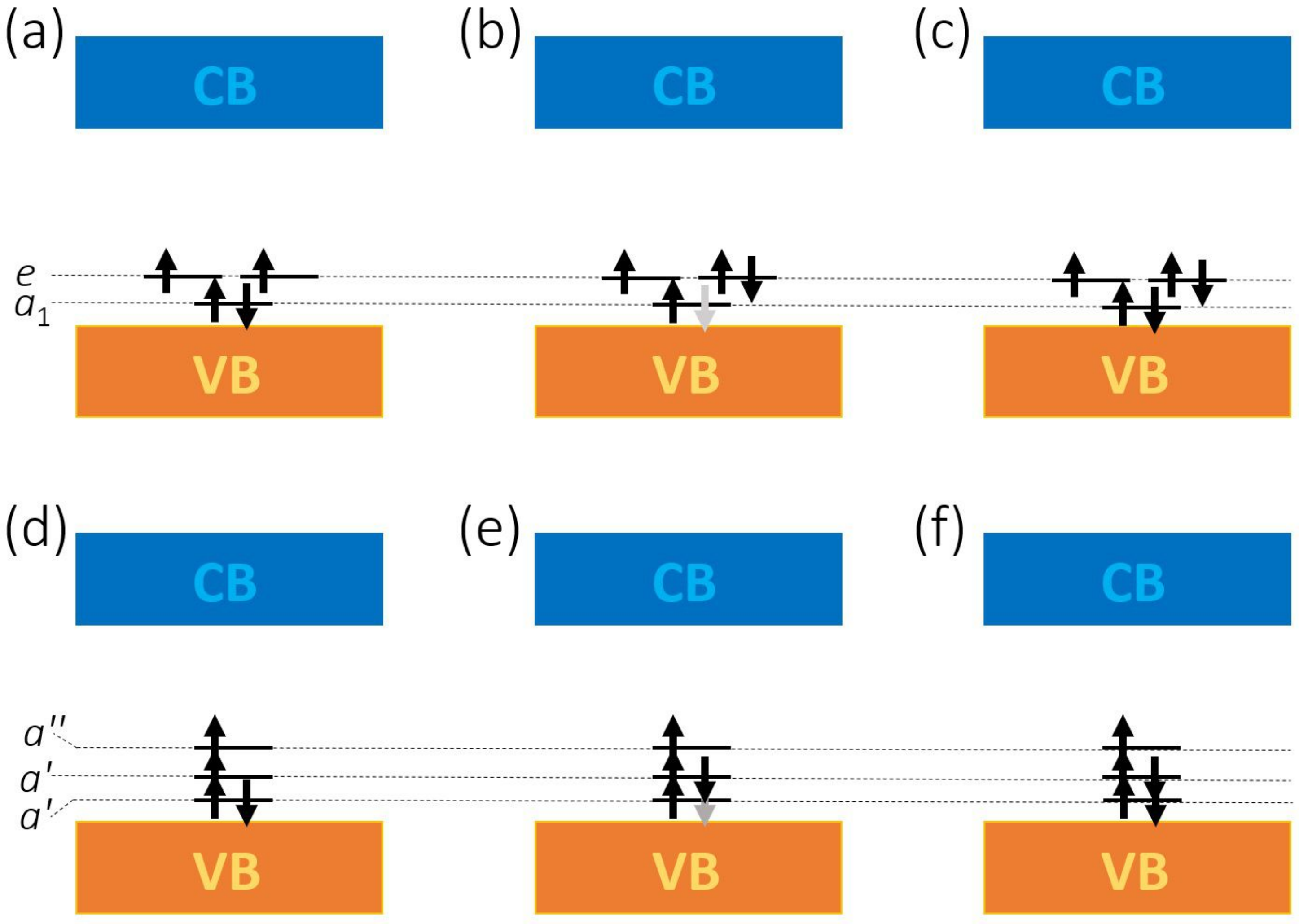}
\caption{Electronic structure of the [(a), (d)] ground and [(b), (e)] excited state of V$_\text{C}$V$_\text{Si}^0$ defects and [(c), (f)] ground state of V$_\text{C}$V$_\text{Si}^-$ defects. Figures (a)-(c) corresponds to the axial defects exhibitng C$_\text{3v}$ symmetry and those of (d)-(f) show the electronic structures of the basal defect configurations possessing C$_\text{1h}$ symmetry. Black arrows represent electrons while gray arrows stand for the reminiscent holes after optical excitation. Valence band (VB) and conduction band (CB) of 4H SiC are also indicated along with the characters of the corresponding Kohn-Sham states. }
\vspace{0 pt}
\label{fig:elstruct}
\end{figure}

In order to obtain the neutral PL spectra for all defect configurations, we calculated the neutral ground and optically accessible excited states by DFT. Excited state calculations were carried out by applying the $\Delta$SCF method~\cite{GaliPRL2009}. In addition, we calculated the ground state of V$_\text{C}$V$_\text{Si}^-$ for the V$_\text{C}$V$_\text{Si}^-$ $\rightarrow$ V$_\text{C}$V$_\text{Si}^0$ reionization spectra. Nuclear coordinates of the global energy minimum in the initial and final electronic states, i.e., the fully relaxed geometries were obtained by minimizing the quantum mechanical forces between the ions falling below the threshold of 0.01~eV/\AA. The corresponding electronic structure for all V$_\text{C}$V$_\text{Si}^0$ defect configurations in the neutral ground/excited state and that for V$_\text{C}$V$_\text{Si}^-$ in the ground state is shown in Fig.~\ref{fig:elstruct}. Here we note that both excited state of V$_\text{C}$V$_\text{Si}^0$ and ground state of V$_\text{C}$V$_\text{Si}^-$ introduce a degenerate $e$ KS state to the band gap occupied by three electrons giving rise to the so-called dynamical Jahn-Teller (JT) effect~\cite{Bersuker2006}. Accordingly, the symmetry will be reduced from C$_\text{3v}$ to C$_\text{1h}$ (even for axial configurations) by coupling to phonons in order to split the corresponding $e$ level removing the degeneracy in total energy. In calculation of the corresponding spectra we use static JT distorted geometries exhibiting the lowest total energy.

All investigated V$_\text{C}$V$_\text{Si}$ defect configurations were modelled in a 576-atom 4H supercell. In order to reach sufficient accuracy in the total energies, we employed $\Gamma$ – point centered 2~$\times$~2~$\times$~2 $k$-point mesh for all ground and excited state calculations. Here we note that only $\Gamma$-point sampling of the Brillouin-zone fails to achieve sufficient accuracy manifesting in the incorrect order in the total energies of the different V$_\text{C}$V$_\text{Si}$ configurations as pointed out in Ref.~\onlinecite{MagnussonPRB2018}. For the electronic structure calculations we employed the HSE06 range-separated hybrid functional~\cite{HeydJCP2003}. Kohn-Sham (KS) wavefunctions were expanded in plane wave basis set with the cutoff energy of 420~eV. Only the valence electrons were treated explicitly, the core-electrons with ionic potentials were considered in the framework of projector augmented wave (PAW) method~\cite{BlochlPRB1994} as implemented in the Vienna Ab-Initio Simulation Package (VASP)~\cite{KressePRB1996}.

For the vibrational modes, we calculated the corresponding dynamical matrix containing the second order derivatives of the total energy by means of the Perdew-Burke-Ernzerhof (PBE)~\cite{PerdewPRL1996} functional. We also determined the corresponding eigenvectors. Here we note that the calculated dynamical matrix and hence the HR and DW factors depend on the supercell size where the former may be a bit underestimated because of neglect of long wavelength phonons.

Localization of the phonon modes can be addressed by the inverse participation ratio (IPR)~\cite{AlkauskasNJP2014, Bell1970JOP} defined as
\begin{equation}
\text{IPR}_k = \frac{N\sum_i\mathbf{u}_i^4}{\big(\sum\mathbf{u}_i^2\big)^2},
\label{eq:IPR}
\end{equation}
where $\mathbf{u}_i$ is the vector displacement amplitude of the $i$th atom in the $k$th phonon mode and $N$ is the number of atoms in the defective supercell that is $N$=574 in our case. Consequently, the IPR falls in the region of $[1, N]$ and it is equal to the number of atoms vibrating in the certain phonon mode.

\section{\label{sec:ResDisc}Results and Discussion}

In Secs.~\ref{subsec:HRPL}~-~\ref{subsec:FreeE} we provide our numerical results for the V$_\text{C}$V$_\text{Si}^0$ PL spectra and the temperature dependent reionization spectra for the V$_\text{C}$V$_\text{Si}^- \rightarrow$ V$_\text{C}$V$_\text{Si}^0$ transition for all V$_\text{C}$V$_\text{Si}$ defect configurations. We also discuss the charge state control and simultaneous qubit manipulation of V$_\text{C}$V$_\text{Si}^0$ in 4H SiC with comparing our novel experimental and simulation results in Sec.~\ref{subsec:Exp}.

\subsection{\label{subsec:HRPL}PL spectra of V$_\text{C}$V$_\text{Si}^0$ configurations}

In order to verify the methodology for calculating the PL spectra~\cite{GaliPRL2009, GaliNatComm2016}, we first compare the total PL spectrum (i.e., including all contributions from the individual V$_\text{C}$V$_\text{Si}$ configurations) obtained from theory with the non-quenched V$_\text{C}$V$_\text{Si}^0$ spectrum reported in Sec.~\ref{subsec:Exp}. The non-quenched PL spectra were recorded at $T=3.5$~K with using two different excitation wavelengths, 976~nm and 1053~nm (c.f., Fig.~\ref{fig:expfig2}) along with a 532-nm repump laser in order to maintain the neutral charge state (i.e., to avoid quenching). We note that our simulations correspond to the $T=0$~K condition which is close to the experimental one.
Excitation energies can be taken into account in the calculations by multiplying the obtained numerical spectra with the optical absorption cross-sections ($\sigma$) of the individual V$_\text{C}$V$_\text{Si}$ configurations, respectively, as $L^\text{EM}(E_\text{ex}, \hbar\omega) = \sigma(E_\text{ex})L^\text{EM}(\hbar\omega)$, where $L^\text{EM}(E_\text{ex}, \hbar\omega)$ denotes the PL lineshape upon the excitation energy of $E_\text{ex}$. In our calculations, we identified $\sigma(E_\text{ex})$ as the transition probability at $E_\text{ex}$ in the absorption PDF, i.e. $\sigma(E_\text{ex}) = \text{PDF}_\text{abs}(E_\text{ex})$. We report the corresponding values in Table~\ref{tab:crosssec}.

\begin{table}[b]
\begin{ruledtabular}
\caption {Absorption cross-sections ($\sigma_\text{th}$) and ratios of the ZPL intensities ($R_\text{th}$) in the numerical PL spectrum for the individual V$_\text{C}$V$_\text{Si}^0$ configurations. We also report the ZPL ratios for the experimental non-quenched spectra recorded at $T=3.5$~K ($R_\text{exp}$). We show values for both 976-nm and 1053-nm excitations. }
\label{tab:crosssec}
\center
\begin{tabular}{@{}ccccccc@{}}
\multirow{2}{*}{Config.}
& \multicolumn{2}{c}{$\sigma_\text{th}$}
& \multicolumn{2}{c}{$R_\text{th}$}
& \multicolumn{2}{c}{$R_\text{exp}$}\\
 & 976 nm & 1053 nm & 976 nm & 1053 nm & 976 nm & 1053 nm\\
\hline
$hh$ & 0.0015 & 0.0013 & 1.00 & 1.00 & 1.00 & 1.00 \\
$kk$ & 0.0018 & 0.0017 & 1.26 & 1.33 & 1.16 & 1.33 \\
$hk$ & 0.0021 & 0.0019 & 1.42 & 1.42 & 2.62 & 2.55\\
$kh$ & 0.0022 & 0.0015 & 1.53 & 1.18 & 3.90 & 1.75\\
\end{tabular}
\end{ruledtabular}
\end{table}

\begin{figure*}[t]
\includegraphics[width=1\textwidth]{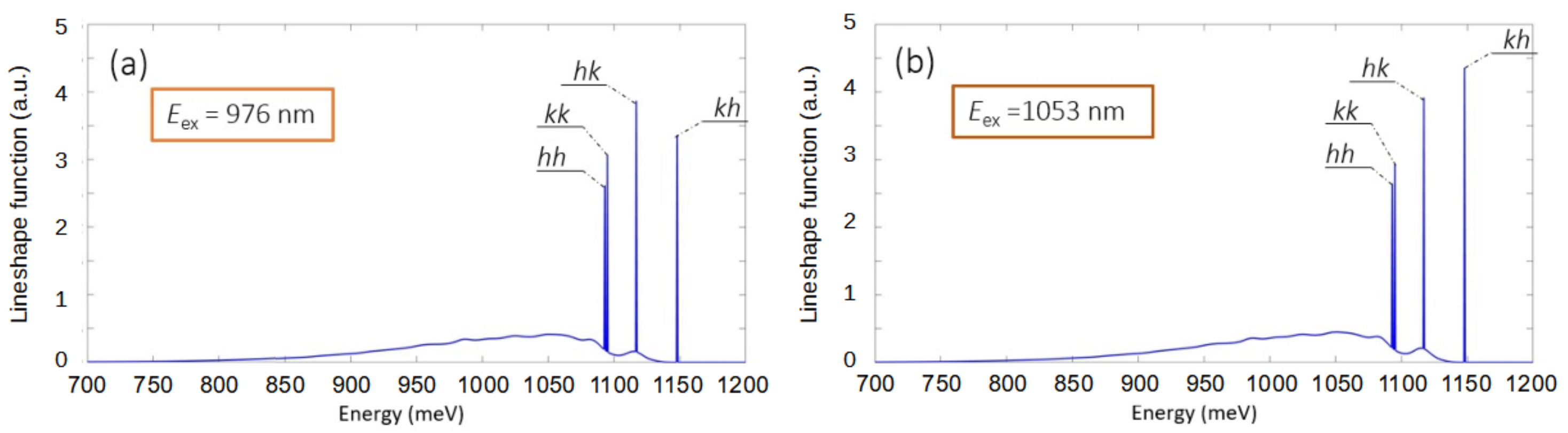}
\caption{Calculated PL spectra of divacancies in 4H SiC upon (a) 976-nm and (b) 1053-nm excitation wavelengths.}
\label{fig:PLthexp}
\end{figure*}

The total numerical PL emission spectra (cf. Fig.~\ref{fig:PLthexp}) upon 976-nm and 1053-nm excitation are then obtained by adding up the PL spectra obtained for the individual V$_\text{C}$V$_\text{Si}$ configurations weighted by the relative strength of absorption probability ($\sigma_\text{th}$) at those excitation wavelengths. Numerical and experimental emission spectra can be well-compared via the ZPL intensities. To this end, we listed the ratios of the numerical and experimental ZPL intesities in Table~\ref{tab:crosssec}. We find that the trend in the ZPL intensity ratios agrees well; however, there are differences between the absolute values. This discrepancy may arise from the fact that, while polarization properties manifest in the experimental spectrum (i.e., the experimental spectrum is the sum of the individual polarization contributions parallel and perpendicular to the $c$-axis), all the emitted photons are considered in the theoretical spectrum. This implies that optical alignment of the PL measurement plays a key role in the recorded lineshape. In order to allow detection of both polarizations simultaneously, the experimental PL spectrum was registered through the edge of the sample as given in Sec.~\ref{subsec:Exp}. Thus, the origin of the discrepancy may be caused by other processes that are not considered in our model.

\begin{figure*}[t]
\includegraphics[width=1\textwidth]{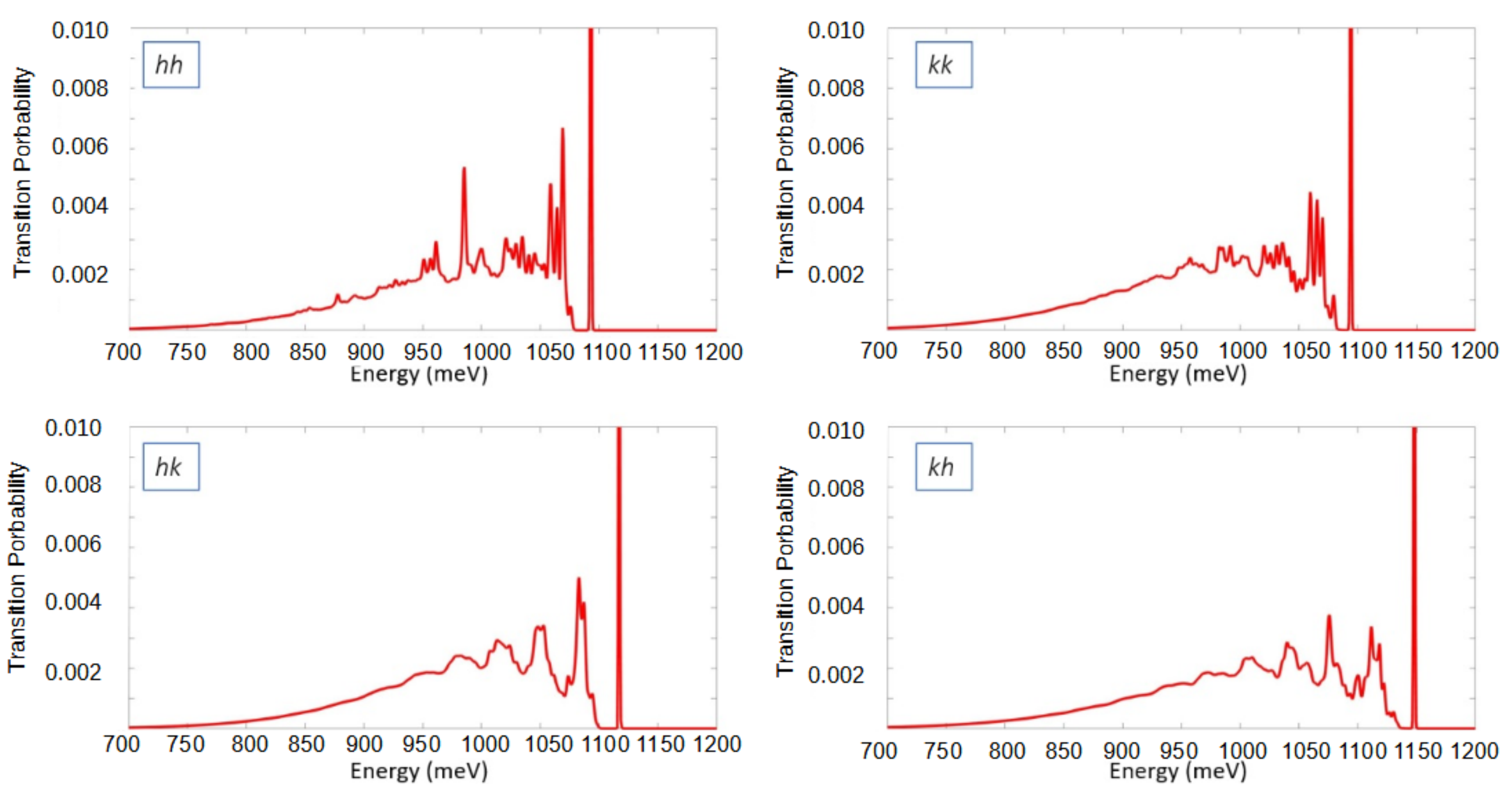}
\caption{Calculated Huang-Rhys emission spectrum for the optical transition of V$_\text{Si}$V$_\text{C}^0$ for all defect configurations (see textbox) assuming resonant excitation. All the spectra are normalized, i.e., the area under the curves is one, hence, the $y$-axis represents the electronic transition probability from the optically accessible excited state to the ground state.}
\label{fig:PL0}
\end{figure*}

\begin{table}[t]
\begin{ruledtabular}
\caption{Calculated and experimental (Ref.~\onlinecite{Falk2013NatCom}) ZPLs for all V$_\text{C}$V$_\text{Si}^0$ defect configurations in 4H SiC. We apply the notations for the centers observed in experiments from Ref.~\onlinecite{KoehlNature2011}. The calculated Huang-Rhys ($S$) and Debye-Waller ($w$) factors are also listed.}
\label{tab:HRDW}
\begin{tabular}{@{}ccccc@{}}
\multicolumn{1}{c}{Config. (Center)}
&\multicolumn{1}{c}{$E_\text{ZPL}^{\text{exp}}$ (eV)}
&\multicolumn{1}{c}{$E_\text{ZPL}^{\text{calc}}$ (eV)}
&\multicolumn{1}{c}{$S$}
&\multicolumn{1}{c}{$w$}\\
\hline
$hh$ (PL1) & 1.095 & 1.142 & 2.856 & 0.057 \\
$kk$ (PL2) & 1.096 & 1.151 & 3.380 & 0.034 \\
$hk$ (PL3) & 1.119 & 1.167 & 3.428 & 0.032 \\
$kh$ (PL4) & 1.150 & 1.187 & 3.540 & 0.029 \\
\end{tabular}
\end{ruledtabular}
\end{table}

\begin{figure*}[t]
\includegraphics[width=1\textwidth]{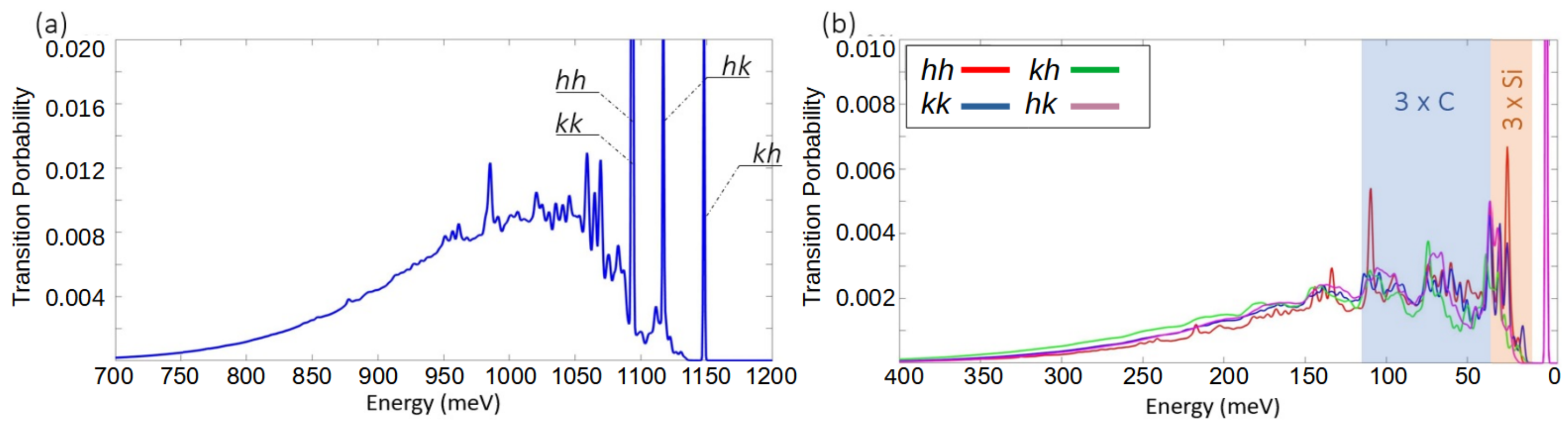}
\caption{(a) Probability density function (PDF) of the V$_\text{C}$V$_\text{Si}^0$ defects in 4H SiC related to (a) total and (b) individual PL spectra. The total spectrum is generated by adding up the individual PDFs of V$_\text{C}$V$_\text{Si}^0$ configurations. ZPL energies are shifted to $0$~meV, in order to read the phonon energy. Localization of phonon modes is indicated by shaded area in (b) caused by the three Si atoms around the V$_\text{C}$ (yellow shaded area) and by the three  
 C atoms around V$_\text{Si}$  (blue shaded area).}
\label{fig:fullPL}
\end{figure*}

The calculated HR emission PDFs of V$_\text{C}$V$_\text{Si}^0$ configurations are shown in Fig.~\ref{fig:PL0}. ZPLs in all spectra are shifted to the experimental ZPL positions as reported in Table~\ref{tab:HRDW}. We note that the fluorescence spectra of V$_\text{C}$V$_\text{Si}^0$ $hh$ and $hk$ configurations have recently been computed in Ref.~\onlinecite{HashemiPRB2021} reporting similar lineshapes. Here we provide the PL lineshape for each configuration with high resolution in the PSB for further analysis. IPR analyis of the phonons shows that phonon modes at $\approx17-35$~meV are mostly localized on the three Si atoms neighboring the V$_\text{C}$ with the IPR of $\approx1-5$ while phonon modes at $\approx35-115$~meV region are mostly localized on the three C atoms around the V$_\text{Si}$ with a similar IPR range [Fig.~\ref{fig:fullPL}(b)]. The sum of the individual PL spectra is shown in Fig.~\ref{fig:fullPL}(a). 

The corresponding HR and DW factors along with the experimental and calculated ZPLs are listed in Table~\ref{tab:HRDW}. The lowest ZPL energy corresponds to V$_\text{C}$V$_\text{Si}^0(hh)$ along with the smallest HR factor and thus the largest DW factor. Here we note that the exceptionally large DW factor of the V$_\text{C}$V$_\text{Si}^0(hh)$ was also found in a previous study~\cite{HashemiPRB2021}. Interestingly, ZPL energy of V$_\text{C}$V$_\text{Si}^0(kk)$ is extremely close to that of V$_\text{C}$V$_\text{Si}^0(hh)$, i.e., the difference between the experimental values is only 1~meV ($\approx0.1$~\%). On the other hand, V$_\text{C}$V$_\text{Si}^0(kk)$ exhibits by $\approx19$~\% larger HR factor and $\approx40$~\% lower DW factor than those of V$_\text{C}$V$_\text{Si}^0(hh)$. We note that the DW factor of V$_\text{C}$V$_\text{Si}^0(hh)$ is close to that of V$_\text{C}$V$_\text{Si}^0$ in the cubic 3C SiC (0.07)~\cite{Csore2022}. We already showed a counterintuitive issue on the example of vanadium defect in 4H SiC~\cite{SpindlbergerPRA2019} that the hexagonal site in 4H SiC has a cubic-like environment in the second and third neighbor shells unlike the quasicubic site. Therefore, $hh$ configuration of divacancy in 4H SiC indeed shows similar properties to those of the divacancy in 3C SiC. The largest ZPL energy and HR factor along with the lowest DW factor occur for V$_\text{C}$V$_\text{Si}^0(kh)$ in 4H SiC.

\subsection{\label{subsec:HRre}Temperature dependent reionization spectra from phonon-assisted transitions}

We conclude that the HR theory V$_\text{C}$V$_\text{Si}^0$ reproduces well the experimental spectrum, therefore we can apply this theory for understanding the temperature dependent reionization process. Ionization energy of point defects can be approximated by their charge transition level which is the position of the Fermi-level in the band gap for which the formation energies of the defect in the two charge states are equal. We showed in our previous study~\cite{MagnussonPRB2018} that the energy separation between the $(+/0)$ level and the valence band maximum (VBM) is at $E_{+/0}^\text{VBM} \approx 1.1$~eV, which is smaller than that between the $(0/-)$ level and the CBM at $E_{-/0}^\text{CBM}\approx1.2-1.3$~eV for all V$_\text{C}$V$_\text{Si}$ configurations in 4H SiC at $T=0$~K. Therefore we concluded that the dark state of the divacancy is V$_\text{C}$V$_\text{Si}^-$, i.e., the singly negative charge state of the defect, into which V$_\text{C}$V$_\text{Si}^0$ is converted upon photoexcitation below the reionization threshold energy because the photon energy of the applied laser used to illuminate V$_\text{C}$V$_\text{Si}^0$ ($>1.1$~eV) could immediately convert V$_\text{C}$V$_\text{Si}^+$ into V$_\text{C}$V$_\text{Si}^0$ with ejecting a hole in the valence band. The experimental threshold energies recorded at cryogenic temperatures and the calculated charge transition levels are listed in Table~\ref{tab:divacTHR}.

\begin{table}[b]
\begin{ruledtabular}
\caption{Experimental threshold energies of photoionization recorded at cryogenic temperatures (Ref.~\onlinecite{MagnussonPRB2018}) which is approximated by the $(0/-)$ charge transition levels for all V$_\text{C}$V$_\text{Si}$ configurations referenced to the CBM. The corresponding HR ($S$) and DW ($w$) factors are also provided which are related to the phonon assisted ionization process.}
\label{tab:divacTHR}
\begin{tabular}{@{}cccccc@{}}
{Center}
&{Config.}
&\multicolumn{1}{c}{$E_\text{rei}$ (eV)} 
&\multicolumn{1}{c}{$E^\text{CBM}_{0/-}$ (eV)}
& {$S$}
& {$w$}\\
\hline
PL1 & $hh$ & 1.310 & 1.245 & 2.691 & 0.068 \\
PL2 & $kk$ & 1.310 & 1.209 & 3.038 & 0.048 \\
PL3 & $hk$ & 1.321 & 1.307 & 2.490 & 0.083\\
PL4 & $kh$ & 1.281 & 1.174 & 3.267 & 0.038 \\
\end{tabular}
\end{ruledtabular}
\end{table}

The reionization threshold energies can depend on the temperature that we divided to three contributors: (i) phonon-assisted photoionization, (ii) entropy contribution to the charge transition levels, (iii) temperature shifts of the band edges of the host crystal. 

\begin{figure*}[t]
\includegraphics[width=0.95\textwidth]{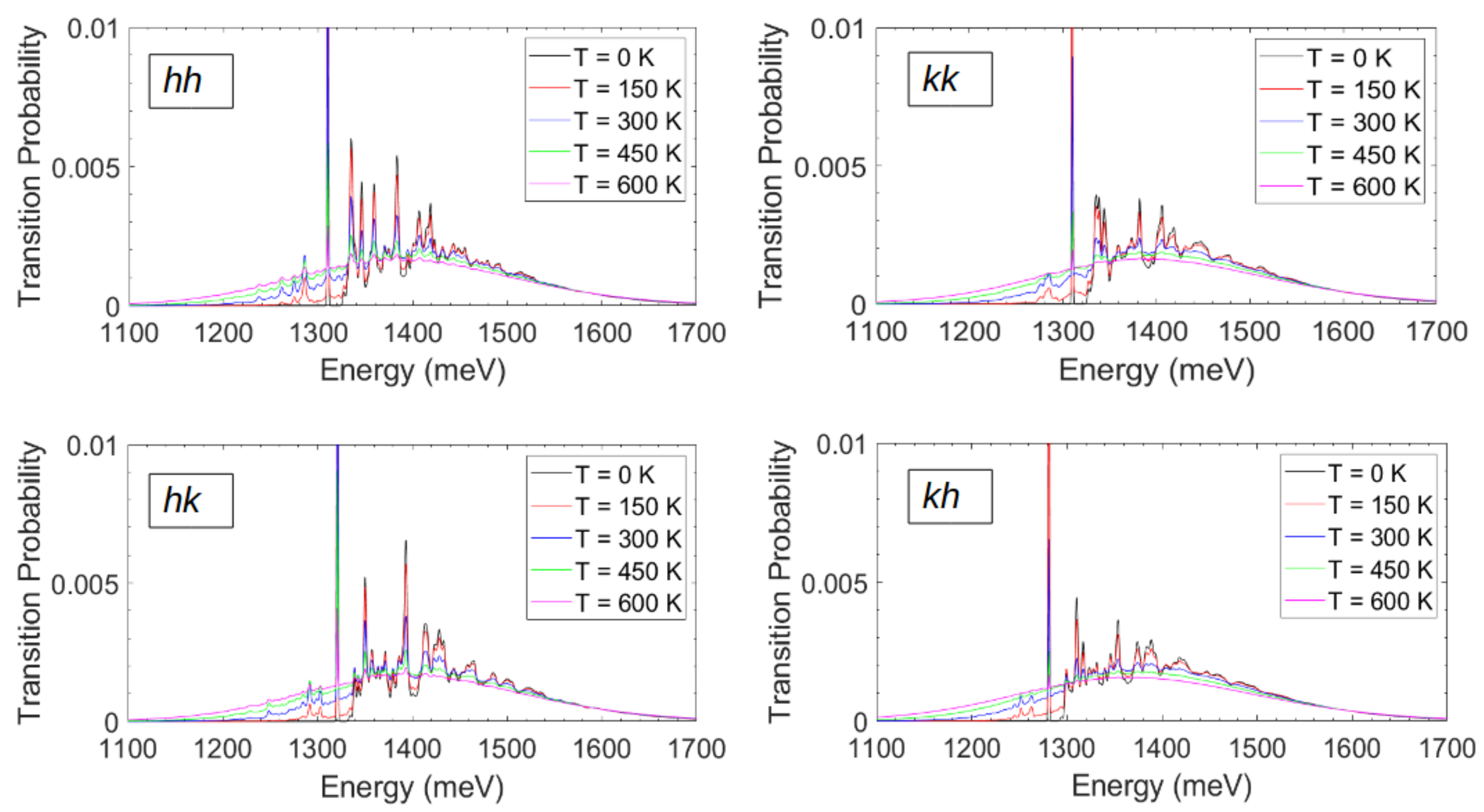}
\caption{Calculated temperature dependent transition probability density functions for HR absorption spectrum of V$_\text{C}$V$_\text{Si}^-$ $\rightarrow$ V$_\text{C}$V$_\text{Si}^0$ transition for the four configurations as indicated in the text boxes 
at $0$~K (black lines), $150$~K (red lines), $300$~K (blue lines), $450$~K (green lines) and $600$~K (magenta lines). The ZPL energies - associated with the reionization energies at cryogenic temperature - were shifted to the experimental data recorded at cryogenic temperatures (Ref.~\onlinecite{MagnussonPRB2018} as listed in Table~\ref{tab:divacTHR}).}
\label{fig:HRreion}
\end{figure*}

First, we consider here the phonon-assisted photoionization process. As we are interested in the threshold energy, we consider only the CBM and the phonon-assisted optical transition to CBM which is technically well represesented in our simulation in terms of folding the M-point of the Brillouin-zone to the $\Gamma$-point of the supercell. The calculated reionization threshold energy, i.e., the $E^\text{CBM}_{0/-}$ in the particular case of divacancies in 4H SiC, can be considered as an absorption of the electron from the highest occupied defect level by the CBM states which corresponds to a no-phonon process. At elevated temperatures, the vibronic excitated states of V$_\text{C}$V$_\text{Si}^-$ may be occupied which effectively lowers the threshold energy to reach CBM compared to that of the no-phonon process. This is analogous to the ZPL energy vs.\ phonon sideband related energies in the absorption process, so HR absorption spectrum calculations were applied for all V$_\text{C}$V$_\text{Si}$ configurations as shown in Fig.~\ref{fig:HRreion}. We used the V$_\text{C}$V$_\text{Si}^0$ phonons for calculating the phonon sideband of the reionization spectra as we did for the PL spectra (see Sec.~\ref{subsec:HRPL}). 

In Fig.~\ref{fig:HRreion}, the ZPL peak corresponds to the no-phonon ionization energies and coincides with the adiabatic $E^\text{CBM}_{0/-}$ charge transition levels. By raising the temperature, peaks and broad features appear at lower energies than the ZPL energy which are responsible to lower the reionization threshold energies. There are some subtle differences between in the low-energy PSB spectrum of the different configurations, e.g., sharp features are visible for $hh$ and $hk$ configurations at $T=600$~K, while those for $kk$ and $kh$ configurations almost completely disappear. This implies a stronger electron-phonon coupling in the reionization process for configurations with V$_\text{C}(h)$ which are associated with the vibration of the three nearest neighbor Si atoms. The HR and DW factors of the absorption (reionization) spectra for each configuration are listed in Table~\ref{tab:divacTHR} which characterize the general strength of the electron-phonon coupling associated with the optical transition.   


The calculation of the phonon-assisted reionization process can be expressed as $E_\text{rei} = E_\text{ex} \pm E_\text{ph}$, where $E_\text{ex}$ represents the applied photoexcitation energy and $E_\text{ph}$ stands for the energy of the absorbed ($-$) or emitted ($+$) phonon. The former process will reduce the threshold energy for reionization. At croygenic temperatures close to $T=0$~K, the no-phonon process is dominating for the threshold energy of reionization, i.e., $E_\text{ex}^\text{min} = E_\text{rei}$ as listed in Table~\ref{tab:divacTHR}. At elevated temperatures, Fig.~\ref{fig:HRreion} clearly shows that, indeed, the threshold energy for reionization may be reduced but the absorption cross-section has an exponential tail in the low-energy PSB of the optical transition. However, it is difficult to define the exact condition at which absorption cross-section the reionization process is observable in experiments. In order to quantify the theoretical description of this process, we setup a condition in which we assume that the reionization process is definitely observable when the total reionization cross-section mediated by the phonons at a given temperature is at least as large as the integrated reionization cross-section associated with the half of the DW factor of the reionization spectrum which corresponds to the reionization cross-section at cryogenic (close to $T=0$~K) temperatures.

Since the calculated reionization spectra are normalized (those are PDFs), exciting with the energy of $E_\text{ex}$ results in the total transition probability , i.e., the cumulative transition probability (CTP) of
\begin{equation}
\text{CTP}(E_\text{ex}) = \int_0^{E_{\text{ex}}} \text{PDF}(\varepsilon)\mathrm{d}\varepsilon,
\label{eq:CTP}
\end{equation}
where $\varepsilon$ is the variable for the excitation energy in the reionization spectra. 
At cryogenic temperatures, we assume that CTP($E_\text{rei} = E_\text{ZPL}$) = $w/2$ with assuming symmetrical broadening of the ZPL peak of the reionization spectrum. By increasing the temperature, the phonon excited states of the negative divacancy are occupied according to the Bose-Einstein function, thus the effective reionization energy decreases which results in a lower threshold energy of photoionization. In other words, the phonon-assisted photoionization results in PDF$(\varepsilon)>0$ for $\varepsilon < E_\text{ZPL}$ energy. One can find the threshold reionization energy at a given temperature for each configuration for which the reionization probability is equal to that of at $T = 0$~K, i.e., PDF$(\varepsilon) = w/2$ is fulfilled. 
\begin{figure*}[t]
\includegraphics[width=0.95\textwidth]{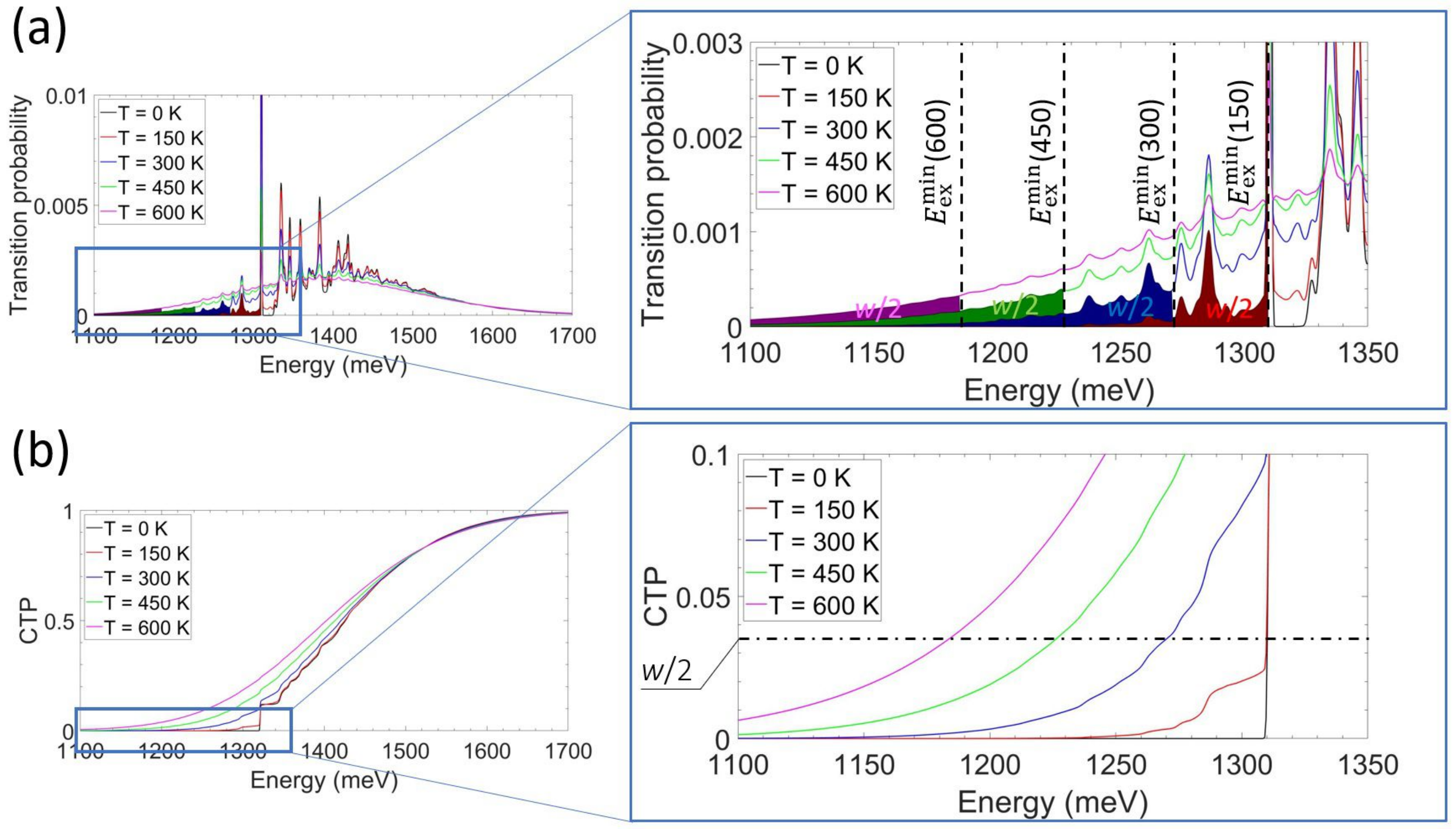}
\caption{(a) Numerical temperature dependent transition probability density functions of the HR absorption (reionization) spectra for the V$_\text{C}$V$_\text{Si}^-(hh)$ $\rightarrow$ V$_\text{C}$V$_\text{Si}^0(hh)$ transition. PSB below the ZPLs given rise at elevated temperatures are enlarged and the corresponding reionization threshold energies ($E_\text{ex}^\text{min}$) are indicated by vertical dashed lines, whereas the cumulative transition probability (CTP) of $w/2$ are represented by shaded areas. (b) CTP for the V$_\text{C}$V$_\text{Si}^-(hh)$ $\rightarrow$ V$_\text{C}$V$_\text{Si}^0(hh)$ transition. Half of the DW factor, i.e., $w/2$ is indicated in the expanded part of the CTP.}
\label{fig:CTPhh}
\end{figure*}

We calculated the CTP($\varepsilon$) at temperatures of $T = 0~\text{K}, 150~\text{K}, 300~\text{K}, 450~\text{K}, 600~\text{K}$ for each divacancy configuration in 4H SiC that we depict for $(hh)$ configuration in  Fig.~\ref{fig:CTPhh}.  CTP($\varepsilon$) can be read out as shaded area under the curves in Fig.~\ref{fig:CTPhh}(a) whereas it is a line plot in Fig.~\ref{fig:CTPhh}(b). CTP($\varepsilon$) at all investigated temperatures is similar to the lineshape of the cumulative distribution function of the Boltzmann distribution. The wavy lineshapes around the ZPL energy visible for low-temperature simulations could be an artifact by neglecting the temperature broadening of the ZPL emission and other size effects of the supercell approach. 

\begin{figure}[t]
\includegraphics[width=0.49\textwidth]{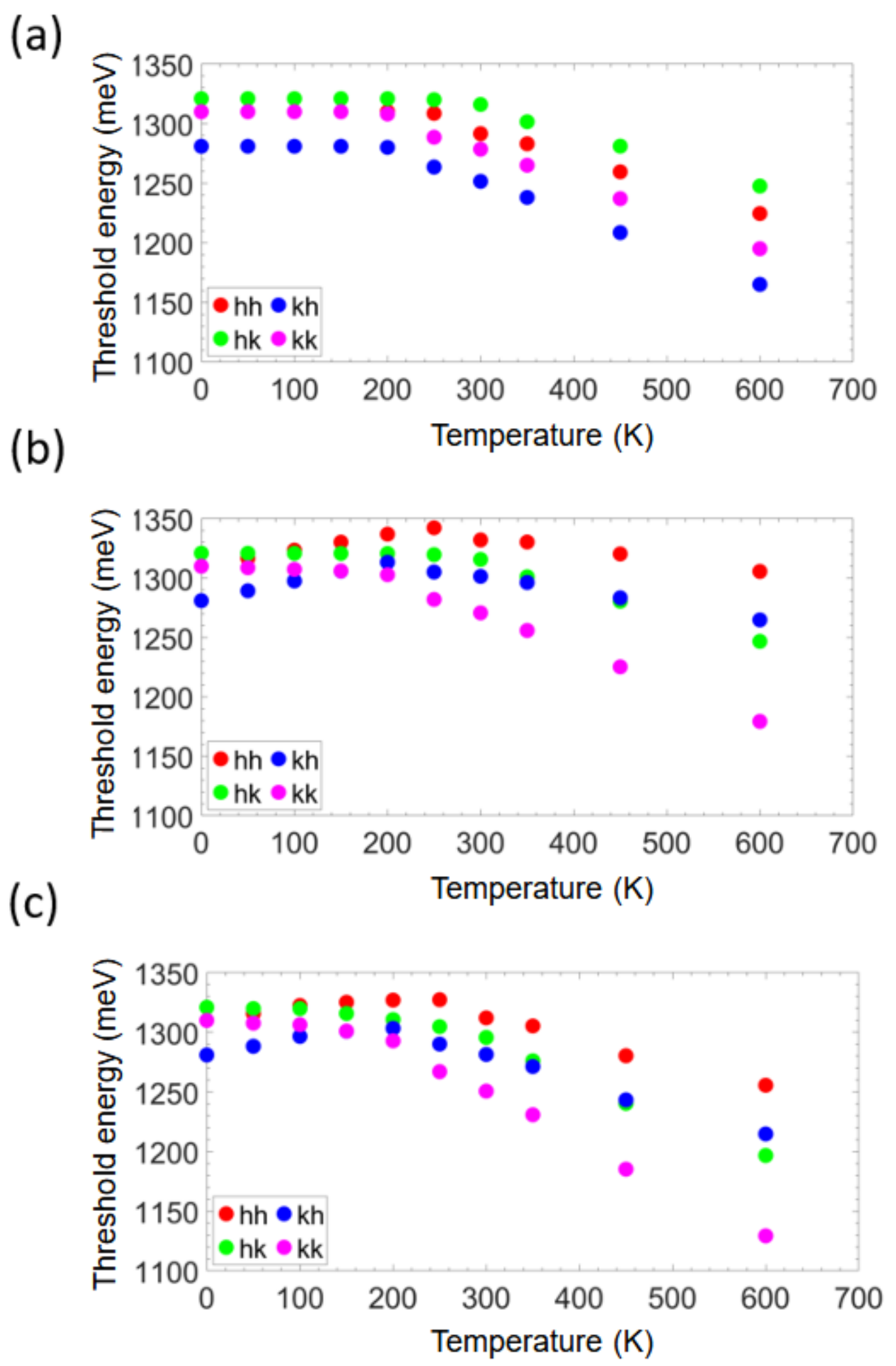}
\caption{Reionization threshold energies [$E_\text{ex}^\text{min}(T)$] (a) without thermal corrections on the charge transition levels, (b) corrected
with $\Delta F(T)$ and (c) corrected further with $\Delta E^\text{CBM}(T)$ as a function of temperature for all V$_\text{C}$V$_\text{Si}$ defect configurations.}
\label{fig:VVTHRT}
\end{figure}

We also indicated $E_\text{ex}^\text{min}$ at each temperature in Fig.~\ref{fig:CTPhh}(a) and depicted for all V$_\text{C}$V$_\text{Si}$ configurations at $T=0\dots 600$~K in $50$~K steps in Fig.~\ref{fig:VVTHRT}(a). This approach results in a plateau up to $T\approx150-200$~K for $E_\text{ex}^\text{min}(T)$, and it then drops for all the V$_\text{C}$V$_\text{Si}$ configurations at higher temperatures. In other words, E$_\text{ph}^\text{min}(T)$ is about zero in our approximation for $T<150-200$~K, and it increases for $T > 150-200$~K. In order to gain further insight to the temperature dependence of the phonon threshold energy, E$_\text{ph}^\text{min}(T)$, we employed the Boltzmann distribution [$f(\varepsilon, T)$] as the low concentration limit of the Bose-Einstein distribution for the phonon bath appearing in the region of [0, $E_\text{rei}$] as
\begin{equation}
f(\varepsilon, T) = \frac{1}{kT} \cdot \text{exp} \Big(\frac{\varepsilon-E_\text{rei}}{kT}\Big),
\label{eq:Boltzmann}
\end{equation}
where $k$=8.617~$\cdot$~10$^{-5}$~eV/K is the Boltzmann-constant. By setting the value of the integral to $w/2$ in the region of [0, $E_\text{rei}$], we arrive at
\begin{equation}
\int_0^{E_\text{ex}^\text{min}} \frac{1}{kT} \cdot \text{exp} \Bigg(\frac{\varepsilon-E_\text{rei}}{kT}\Bigg)\mathrm{d}\varepsilon = \frac{w}{2}
\label{eq:Boltzint}
\end{equation}
By evaluating the integral of Eq.~\ref{eq:Boltzint} and rearranging the equation, we arrive at
\begin{equation}
E_\text{ph}^\text{min}(T) = -\text{ln}\Bigg[\frac{w}{2} + \text{exp}\Bigg(\frac{E_\text{rei}}{kT}\Bigg)\Bigg] \cdot kT \text{.}
\label{eq:THRsol}
\end{equation}
Since the values of $E_\text{rei}$ are around 1300~meV the typical values of $\text{exp}\big(\frac{E_\text{rei}}{kT}\big)$ is about $10^{-12}$ at the given temperature range, and hence negligible next to the calculated DW values of $w/2 \approx 0.03-0.09$  (see Table~\ref{tab:divacTHR}). Finally, $E_\text{ph}^\text{min}(T)$ can be approximated as 
\begin{equation}
E_\text{ph}^\text{min}(T) \approx -\text{ln}\Bigg(\frac{w}{2}\Bigg) \cdot kT
\label{eq:THRapprox}
\end{equation}
yielding linear characteristics for $E_\text{ph}^\text{min}(T)$ with the slope of $m = -\text{ln}(w/2) \cdot k$. Furthermore, since $w/2 < 1$, $E_\text{ph}^\text{min}(T)$ increases with $T$ within this approximation. As a result, reionization threshold energy can be expressed as
\begin{equation}
E_\text{ex}^\text{min}(T) = E_\text{rei} - E_\text{ph}^\text{min}(T) \approx E_\text{rei} + \text{ln}\Bigg(\frac{w}{2}\Bigg) \cdot kT \text{.}
\label{eq:Erei}
\end{equation}
Linear behavior of $E_\text{ex}^\text{min}(T)$ is indeed manifested in our simulations plotted in Fig.~\ref{fig:VVTHRT}(a), for $T > 200$~K. This approach predicts that $E_\text{rei}(T=300~\text{K})$ reaches 1250-1320~meV (990-940~nm) for V$_\text{C}$V$_\text{Si}$ configurations in 4H SiC which is close to the typical photoexcitation energy of the neutral divacancies in 4H SiC. 
We note that Eq.~\ref{eq:Erei} is valid for any point defects and their PL spectrum within the employed approximations by substituting $E_\text{rei}$ with the corresponding ZPL energy of (re)ionization. We emphasize that the actual phonon-assisted reionization may be observed for $\varepsilon$ with PDF$(\varepsilon) < w/2$ depending on the integration time and other factors in the experiments that are discussed in Sec.~\ref{subsec:Exp}. Furthermore, the temperature shifts of the charge transition level and the band edges, i.e., temperature dependence of $E_\text{rei}$, should be also taken into for understanding temperature dependence of the reionization process. 

\subsection{\label{subsec:FreeE} Temperature dependent reionization thershold energies including contributions from entropy and temperature shifts of the conduction band minimum}

Temperature dependence of the charge transition levels of point defects was discussed in Sec.~\ref{subsec:CTLTdep} which depends on the entropy effects in the formation enthalpy of the point defects in their charge states ($\Delta F(T)$) and the electron-phonon renormalization of the band edges. Thermal evolution of the band edges in 4H SiC was already determined in a previous study~\cite{CannucciaPRMat2020}. According to this work, thermal shifts of the VBM and CBM becomes noticeble at around $T=300$~K, where the CBM/VBM is decreased/increased by $\approx5$~meV, respectively, with respect to their $T=0$~K value. In Table~\ref{tab:Ereicorr}, we provide the values for $\Delta F(T)$ and $E^{\text{CBM}}_{-/0}(\text{corr})$ for all V$_\text{C}$V$_\text{Si}$ defect configurations at room-temperature.
\begin{table}[t]
\begin{ruledtabular}
\caption{Free energy correction term ($\Delta F$) to $E^{\text{CBM}}_{-/0}$ and the corrected charge transition levels referenced to the CBM [$E^{\text{CBM}}_{-/0}(\text{corr})$] at room temperature ($T=300$~K). We already included the $\Delta E_\text{CBM}(T)=5$~meV decrease of the CBM in the calculation of $E^{\text{CBM}}_{-/0}(\text{corr})$. }
\label{tab:Ereicorr}
\begin{tabular}{@{}lcccc@{}}
Config. & $hh$ & $kk$ & $hk$ & $kh$\\
\hline
$\Delta F$(eV) & 0.041 & -0.008 & -0.001 & 0.050\\
$E^{\text{CBM}}_{-/0}(\text{corr})$ (eV) & 1.281 & 1.196 & 1.301 & 1.219\\
\end{tabular}
\end{ruledtabular}
\end{table}  

We find that the $\Delta F$ is one order magnitude lower for the $kk$ and $hk$ configurations than those for the $hh$ and $kh$ configurations. Since $F^q(T)$ depends on frequency of the phonon modes (cf.\ Eq.~\ref{eq:freeE}), $\omega_i$, this trend may arise from the stronger distortion between the neutral and negative charge states for the $hh$ and $kh$ configurations than that for the $kk$ and $hk$ configurations. Indeed, the geometry distortion is larger by about 10\% for the $hh$ and $kh$ configurations than that for $kk$ and $hk$ configurations. As a result, correction in $E^{\text{CBM}}_{-/0}$ (cf.\ Table~\ref{tab:divacTHR}) for the $kk$ and $hk$ configurations is tiny and further reduces the reionization threshold energies. On the other hand, the free energy contribution rather inceases the reionization threshold energies for $hh$ and $kh$ configurations and mostly compensates the thermal shift of the band edge [cf. Figs.~\ref{fig:VVTHRT}(b) and (c)]. These results clearly demonstrate that considering only the free energy term in the temperature dependence of the ionization threshold energies can lead to a false result as the contribution from the phonon-assisted ionization process could dominate and even change the sign of the thermal shift in the ionization threshold energy [cf. Fig.~\ref{fig:VVTHRT}(a) and (c)].

\subsection{\label{subsec:Exp} Experimental aspects}

The above consideration shows that the quenching behavior of the divacancy in all configurations is distinctly different at low temperature (say, < 2 K) and at elevated temperatures. While at low temperature there is a distinct threshold for the excitation energy for each divacancy configuration, below which quenching of the corresponding PL is observed but can be recovered using higher-energy repump illumination. Above the threshold the PL of the corresponding divacancy configuration does not quench even if no repump is applied. At higher temperatures, however, there are no distinct thresholds because phonon-assisted photoionization of the VV$^-$ charge state becomes possible, whereas no phonons are available at low temperature for this process. Given a fixed excitation energy $E_\text{ex}$, the minimum energy of a phonon $E_\text{ph}^\text{min}$ that can combine with the incident photon to produce phonon-assisted photoionization of VV$^-$ is given by the energy balance quoted in Sec.~\ref{subsec:HRPL},
\begin{equation}
E_\text{rei} = E_\text{ex} + E_\text{ph}^\text{min}
\label{eq:Erei1}
\end{equation}
Here we assume that $E_\text{ex}<E_\text{rei}$, hence $E_\text{ph}^\text{min}>0$. All phonons with energies $E_\text{ph}>E_\text{ph}^\text{min}$ also contribute to phonon-assisted photoionization, hence the cumulative transition probability CTP($E_\text{ex}$) is given by Eq.~\ref{eq:CTP}.
From practical point of view, it may be desirable to have the possibility for optical charge state control also at elevated temperatures. This can be achieved by using lower excitation energies $E_\text{ex}$, because these provide lower CTP($E_\text{ex}$), hence the quenching behaviour can be preserved also at elevated temperatures. It is clear that if $E_\text{ph}^\text{min} \gg k_\text{B}T$ for given temperature $T$ (the case of lower-energy E$_\text{ex}$) the amount of phonons available for phonon-assisted reionization is negligible and CTP($E_\text{ex}$)~$\approx$~0. Note that $E_\text{ex}$ is bound from below by the condition $E_\text{ex}>E_\text{ex}^\text{ZPL}$, where $E_\text{ex}^\text{ZPL}$ refers to the corresponding divacancy configuration. Our goal here is to examine the concept of phonon-assisted reionization by comparing the quenching properties at higher temperatures for two excitation energies, both below the thresholds, but one of which is close to the threshold while the other is far enough so that $E_\text{ph}^\text{min}= E_\text{rei}-E_\text{ex} \gg k_\text{B}T$.

\begin{figure} [t]
\includegraphics[width=1.0\textwidth]{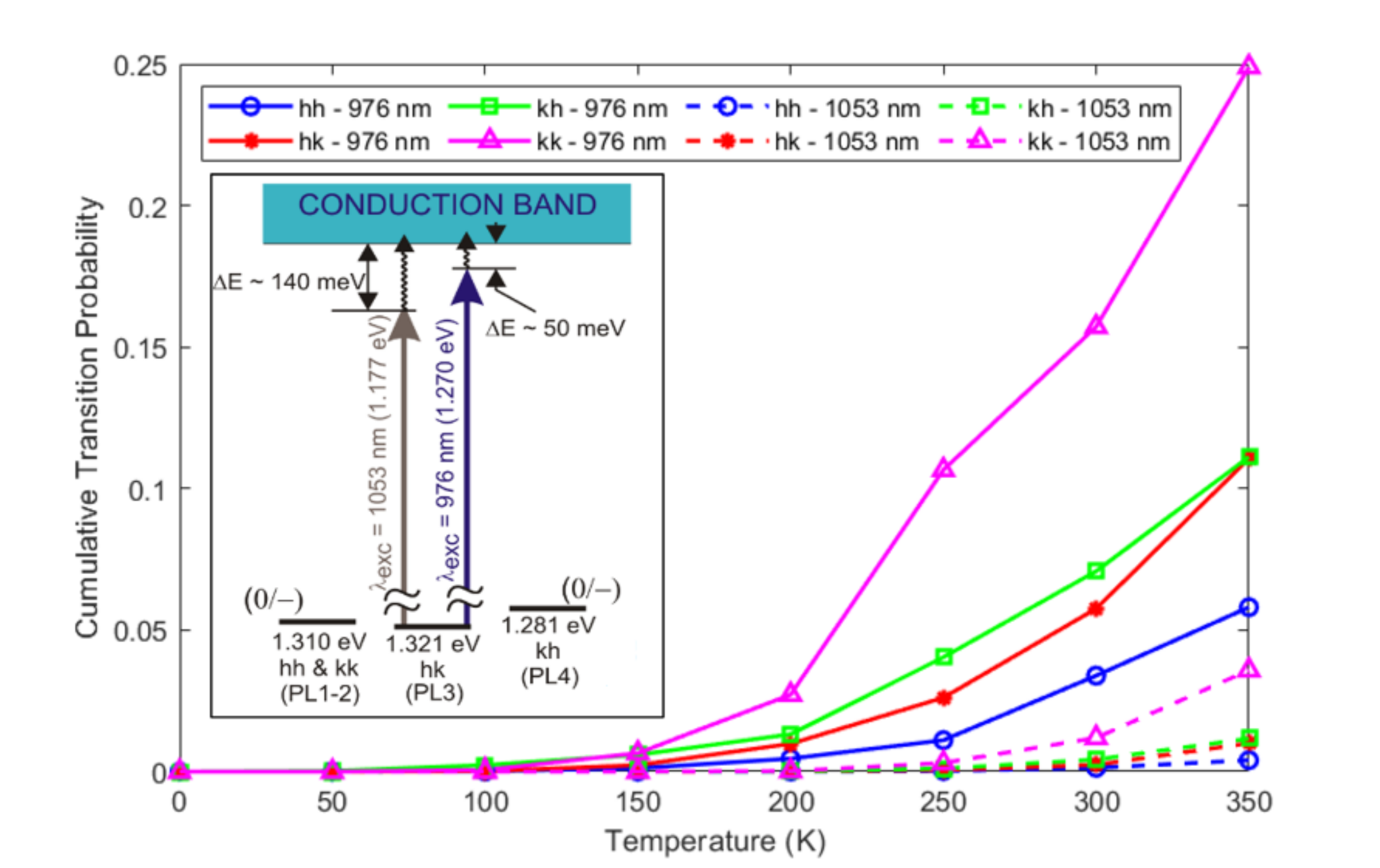}
\caption{Theoretical dependencies of the cumulative transition probability on the temperature for the four divacancy configurations as denoted for each curve taking all the temperature effects into account, and for the two excitation wavelengths used in the experiment. The inset displays an energy diagram of the phonon-assisted photoionization illustrated on the hk divacancy configuration. The $(0/-)$ charge transition levels are labelled with the corresponding configuration and the experimentally determined energy separation from the conduction band. The two bold vertical arrows represent the two experimental laser energies whereas the wavy arrows represent the phonons with minimum energies needed for reionization of the divacancy from negative to neutral charge state.  }
\vspace{0 pt}
\label{fig:expfig1}
\end{figure}

\begin{figure} [b]
\includegraphics[width=0.7\textwidth]{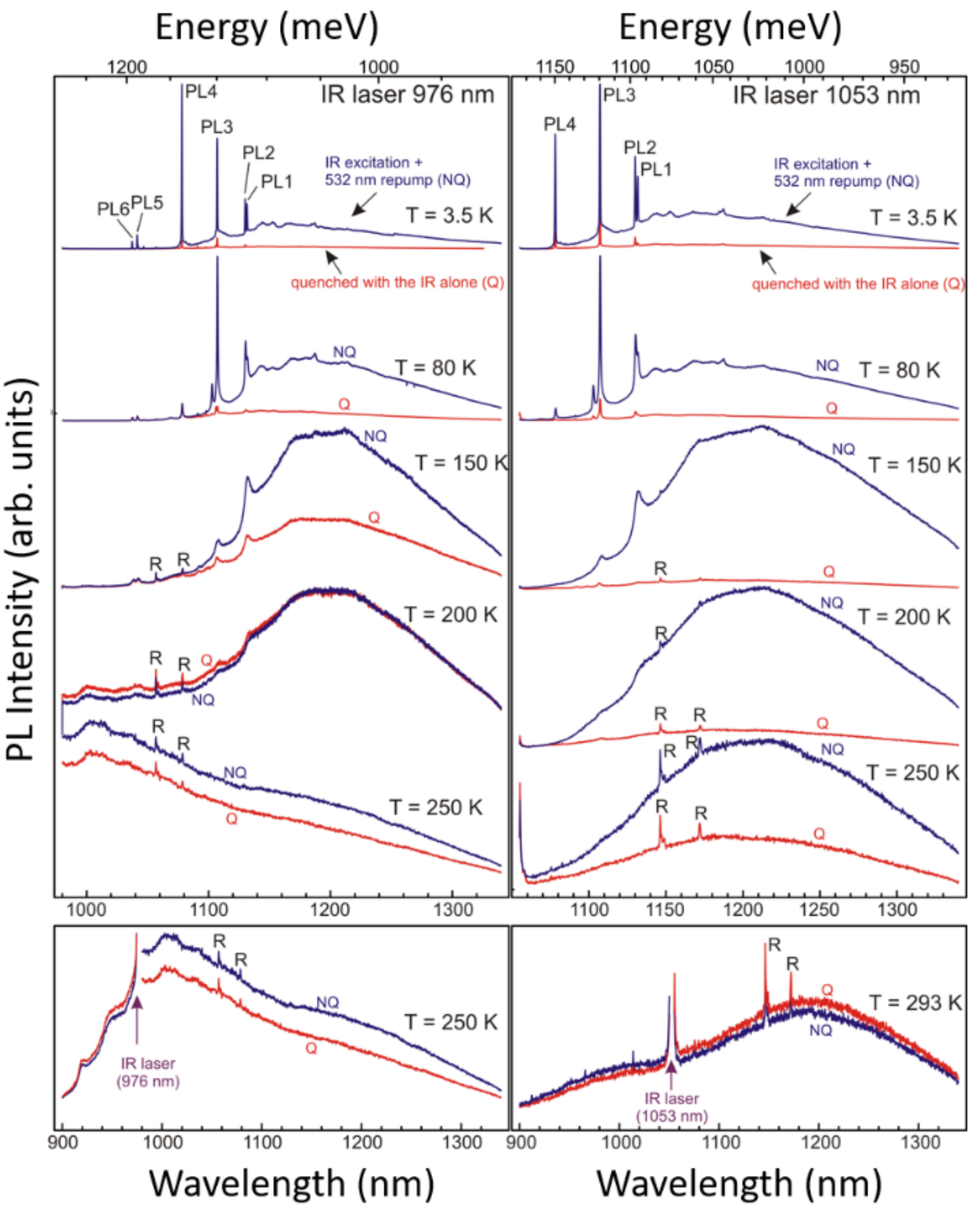}
\caption{Experimental spectra of the divacancy emission obtained at different temperatures with the simultaneous application of the IR laser and the $532$~nm repump laser (denoted NQ, or non-quenched) and with the IR laser alone after a long term quenching ($Q$, or quenched). The panels to the left (right) refer to the $976$~nm ($1053$~nm) IR laser, respectively. The two bottom panels show also the anti-Stokes part of the spectra and illustrate the upconversion in the emission of the silicon vacancy. }
\vspace{0 pt}
\label{fig:expfig2}
\end{figure}

The energy diagram (inset in Fig.~\ref{fig:expfig1}) depicts the two excitation cases for $E_\text{ex}=1.270$~eV ($976$~nm), and $E_\text{ex}=1.177$~eV ($1053$~nm) which have been used in the experiments presented below. Both excitations are below the thresholds for all divacancy configurations, but the former is close to the thresholds (only $\approx11$~meV lower than the threshold for PL4, cf. Table~\ref{tab:HRDW}), whereas the $1.177$~eV excitation is about $100$~meV (up to $140$~meV for the different configurations) below all thresholds. As a consequence, the condition $E_\text{ph}^\text{min} \gg k_\text{B}T$ is fulfilled only for the second excitation at $1.177$~eV, as long as the temperature does not exceed room temperature. The CTPs as a function of temperature for the two used excitations and for the four divacancy configurations are calculated using Eq.~\ref{eq:CTP} and displayed in Fig.~\ref{fig:expfig1} in the temperature range $0–350$~K. Using these dependencies we can make qualitative comparison with the experimental data presented in Fig.~\ref{fig:expfig2} and obtained using these two fixed laser excitations, $976$ and $1053$~nm. The repump laser beam used in these experiments is at $532$~nm directed unfocused on the sample with power density of the order of $1$~mW/cm$^{-2}$. On the other hand, the excitation beam ($976$ or $1053$~nm) is moderately focused on the sample to a spot of $\sim$1~mm diameter, with estimated power density at the sample of the order of several watts per cm$^{-2}$. The experimental conditions are detailed in Ref.~\onlinecite{MagnussonPRB2018}.

We compare now the quenching behaviour of the divacancy emission at different temperatures when the two different excitations are used. The experimental spectra obtained at different temperatures are presented in Fig.~\ref{fig:expfig2}. The spectra with the infrared laser excitation alone ($976$ or $1053$~nm) are obtained after switching off the repump laser and quenching the divacancy PL for about 1/2 hour, and are referred to further as the “quenched” spectra (denoted by $Q$ in Fig.~\ref{fig:expfig2}). The rest of the spectra are with the repump laser applied in addition to the infrared excitation; under these conditions the divacancy PL is stable and the spectra are denoted by NQ for “non-quenched” in Fig.~\ref{fig:expfig2}. 
One can see from the spectra that at liquid He temperature ($3.5$~K) the divacancy PL quenches about $10–30$ times for each of the two excitations, as expected because both excitation energies are below the thresholds for all divacancy configurations. Also, at this temperature essentially no phonons are available for phonon-assisted reionization, hence the accumulation of the divacancy in the negative charge state is irreversible for both IR excitations used. We notice also that in the particular sample used in these experiments the quenched PL never reaches zero intensity, similar to sample 1 of Ref.~\onlinecite{MagnussonPRB2018}. 

At 80~K, however, we notice differences in the quenching behavior of one of the divacancy configurations ($kh$). Namely, with the 976-nm excitation the PL4 line corresponding to the $kh$ configuration shows nearly the same intensity in the “quenched” spectrum as in the “non-quenched”, i.e., it does not quench. We notice here that the relative contribution of the PL4 line in the spectra at $80$~K is strongly reduced compared to the spectra at $3.5$~K, but this behavior is not understood at present. The rest of the lines (PL1-PL3) do quench exhibiting large intensity contrast (similar to that at $3.5$~K) between the “quenched” and “non-quenched” spectra for both excitations. We notice further that the PL4 also quenches in a way similar to that at $3.5$~K if the 1053-nm excitation is used. Referring to the CTP plots in Fig.~\ref{fig:expfig1} we notice that at temperature $T=80$~K only the $kh$ configuration (corresponding to the PL4 line) exhibits CTP$\approx0.0015$ for 976-nm excitation, while for the rest of the divacancy configurations at $976$~nm and all configurations at $1053$~nm the CTP is essentially negligible ($< 3.5\times10^{-4}$) for both excitation wavelengths. Thus, the experimental data for the PL4 line implies that for CTP of only 0.0015 (0.15\% probability of phonon-assisted ionization of V$_\text{C}$V$_\text{Si}^-$) the reionization process outcompetes the accumulation of the divacancy in the negative charge state. Consequently, for CTPs in the low 10$^{-3}$ range one may expect to notice impact of the phonon-assisted reionization on the quenching process. This provides a means to define semi-quantitatively the experimental conditions when the rate of the phonon assisted reionization is comparable with the rate of capture of electrons to neutral divacancies, thus creating negatively-charged ones. We can define that this condition (similar rates of ionization an reionization) is true if the quenched spectra exhibit roughly 50\% of the intensity of the non-quenched ones.

At temperatures higher than $80$~K the contribution of the $kh$ configuration (PL4 line and phonon sideband) in the spectrum can be neglected owing to the abovementioned rapid decrease of the PL4 intensity with increasing temperature. Thus, we investigate the behavior of the PL1-PL3 lines, the phonon sidebands of which coalesce into a broad band extending between $1100$ and $1320$~nm. Note that also the threshold energies of these three configurations ($hk$, $hh$ and $kk$) are rather close according to the energy diagram inset in Fig.~\ref{fig:expfig1}. At $T=150$~K, the contributions from the these three divacancy configurations are still discernible via their zero-phonon lines, although the latter are much broader and weaker than at cryogenic temperature. However, the dominant contribution to the spectrum at this and higher temperatures comes from the PSBs of each configuration, which overlap to form a single band concealing the individual contribution from the different configurations. Therefore, in the following we shall compare the intensities of the broad bands in the “quenched” and “non-quenched” spectra, looking for each infared (IR) excitation for the temperature when the intensity in the quenched spectrum is roughly half of the intensity in the non-quenched one.

We observe from Fig.~\ref{fig:expfig2} that our criterion is fulfilled for the 976-nm excitation at $T=150$~K, when the intensity of the phonon sideband is about 1/2 of the intensity in the non-quenched one. The corresponding CPTs for the 976-nm excitation are $2.3\cdot10^{-3}$, $1.2\cdot10^{-3}$ and $6.5\cdot10^{-3}$ for the $hk$, $hh$, and $kk$ configurations, respectively. These values are in good agreement with our anticipation that the phonon-assisted reionization rate is comparable with that of the divacancy ionization to V$_\text{C}$V$_\text{Si}^-$. Further temperature increase to 200~K completely inhibits the quenching as seen from the corresponding spectrum for 976~nm excitation in Fig.~\ref{fig:expfig2}. On the other hand, the quenching is well pronounced for the 1053~nm excitation at both temperatures, 150~K and 200~K. For this excitation (1053~nm), one notices significant change in the quenching contrast only at 250~K, when the intensity of the “quenched” spectrum is about 30\% of that of the “non-quenched” one. The corresponding theoretical CPTs are $1.2\cdot10^{-4}$, $4.4\cdot10^{-4}$ and $3.1\cdot10^{-3}$ for the $hk$, $hh$, and $kk$ configurations, respectively. These values imply that $kk$ configuration contributes the most or entirely to the “non-quenched” spectrum, and we still expect impact on the quenching properties for such values of the CPT for $hk$ and $hh$ configurations. Thus, the experimental data is in reasonable agreement with the theory and we may conclude that the dominant reason for the loss of the quenching properties above $T=150$~K quoted in Ref.~\onlinecite{MagnussonPRB2018} where 976-nm excitation has been used is the phonon-assisted reionization activated at this and higher temperatures. In contrast, the quenching property disappears completely only at room temperature when the 1053~nm excitation is used. Indeed, the CPT values in these conditions are $2.4\cdot10^{-3}$, $1.3\cdot10^{-3}$ and $1.2\cdot10^{-2}$ for the $hk$, $hh$, and $kk$ configurations, respectively, which are all above the quoted CPT threshold value. 

We notice the appearance of the luminescence of the silicon vacancy (V$_\text{Si}^-$) at elevated temperatures when the 976-nm excitation is used, via the upconversion mechanism discussed in~Ref.~\onlinecite{MagnussonPRB2018}. Indeed, at $T=250$~K the spectrum obtained with 976-nm excitation is dominated by the contribution of V$_\text{Si}^-$, as shown in the bottom left panel of Fig.~\ref{fig:expfig2}. However, the V$_\text{Si}^-$ spectrum can be discerned even below $T=200$~K when 976-nm excitation is used. For instance, the Stokes part of the V$_\text{Si}^-$ luminescence is clearly visible in the spectrum at $T=200$~K (left panels). In contrast, weak contribution from the upconverted V$_\text{Si}^-$ luminescence can be observed with 1053-nm excitation only at room temperature ($293$~K, bottom right panel); at $T=250$~K or lower temperatures its contribution to the spectrum is negligible and the dominant luminescence is that of the divacancy. Since the upconverted V$_\text{Si}^-$ luminescence is associated with increased generation of free electrons~\cite{MagnussonPRB2018}, its appearance serves as an indicator of the rapid increase of the free-electron population with temperature when 976-nm is used, owing to the rapid increase of the CTP with increasing temperature for this excitation. On the contrary, the increase in the CTP (and the electron population) is much smaller with the 1053-nm excitation, so that the contrast between quenched and non-quenched spectra disappears only at $T=293$~K and is associated with the weak appearance of upconverted luminescence from V$_\text{Si}^-$, as expected.

Thus, the experimental data suggests that the phonon-assisted divacancy reionization apparently influences the quenching behavior at temperatures and excitations for which the CTP is of the order of low $10^{-3}$, i.e., well below 1\%. We notice that this value is much smaller than the half of the Debye-Waller factor ($w/2$) used as a reference during the discussion of the probabilities in the previous section. We can understand the fact that low reionization probability (of the order of 0.1\%) is sufficient to counteract the process of accumulation of the divacancy in the negative charge state, if we consider the model describing the quenching dynamics presented in Ref.~\onlinecite{MagnussonPRB2018}. In this simplest model two types of centers are considered: traps capable to emit electrons to the conduction band as a result of photoionization (via the IR excitation) and divacancies capable of capturing electrons. Consequently, the traps can be in two states: with captured electron (capable to emitted an electron to the conduction band via photoionization), and with a missing electron (capable to recapture of free electron). The divacancies in this model are also assumed to have two charge states: neutral (capable of capturing electron) and negatively-charged (with captured electron). The latter state cannot be photoionized using IR excitation below the threshold, hence the accumulation of divacancies in the negative charge state leads to quenching of the luminescence from the neutral charge state.

Naturally, there exist various traps capable of emitting electrons via photoionization, for instance, the nitrogen donors, the carbon vacancy, the silicon vacancy in double (2$-$) and triple (3$-$) negative charge states, just to name a few. The model in Ref.~\onlinecite{MagnussonPRB2018} assumes only one kind of traps which may be thought as having the dominant capture cross section for electrons and the dominant concentration. If these were the nitrogen donors, for instance, then the capture cross-section of ionized donors for electrons is “giant”~\cite{LaxPR1960} owing to the possibility for capture via the excited donor states. On the other hand capturing of electrons occurs to the neutral divacancy which, therefore, is assumed to have much smaller capture cross section for electrons. Indeed, in order to reproduce correctly the shape of the time decay of the divacancy PL the model in Ref.~\onlinecite{MagnussonPRB2018} admits that the ratio of capture cross sections for electrons of the divacancy and the trap is $2\cdot10^{-3}$. In other words, an electron is emitted from traps and recaptured from traps many times before it eventually can be captured by the divacancy. On the other hand, one can assume that the probability for absorption of a photon which is stipulated by the optical-absorption cross-section is similar for the traps and the divacancies. Recapturing of electrons is fast to the traps, but only about 0.1 \% of the total free-electron population are captured to divacancies. Thus, the divacancy population in the negative charge state grows slowly, and this slow process can be counteracted by phonon-assisted reionization with a CTP of about 0.1 \%. 

To conclude this section, we can state that excitations with longer wavelength (lower energy) widen the temperature range in which the quenching phenomenon can be observed. However, in many cases when the silicon vacancy V$_\text{Si}^-$ is also present, excitation with too high photon energies (e.g., close to or even above the threshold of all divacancy configurations) leads at elevated temperatures to upconverted PL of V$_\text{Si}^-$ instead of excitation of the divacancy PL, as illustrated in the low left panel of Fig.~\ref{fig:expfig2}. Thus, if the PL of the divacancy is to be measured at higher temperatures or at room temperature, lower excitation energies are also preferable. It should be noted, however, that it might be possible to stabilize the divacancy PL by Fermi-level tuning even when excitations above the thresholds is applied, but this is a subject of a different study and will not be discussed here.

\section{\label{sec:Concl}Conclusions}
In this paper, we investigated the photoluminescence lineshape of the V$_\text{C}$V$_\text{Si}^0$ qubits in 4H SiC at $T=0$~K by DFT calculations. We obtained the corresponding PL spectra by using the HR theory and found that the corresponding PSB exhibits a rich set of sharp features for each configuration. The sharp features originate from quasilocalized phonon modes. We showed via IPR analysis that phonons are localized on the three neighbor Si atoms around the V$_\text{C}$ in the phonon energy region of $\approx17-35$~meV while those falling in the $\approx35-115$~meV are localized on the three neighbor C atoms around the V$_\text{Si}$. We also calculated the HR and DW factors of the PL spectra yielding the smallest HR factor (and thus the largest DW factor) for V$_\text{C}$V$_\text{Si}^0(hh)$ implying V$_\text{C}$V$_\text{Si}^0(hh)$ to be the most attractive V$_\text{C}$V$_\text{Si}^0$ qubit candidate in terms of extracting the brightest optical transition at the coherent ZPL emission among the divacancy configurations in 4H SiC. These optical properties of V$_\text{C}$V$_\text{Si}^0(hh)$ in 4H SiC are reminescent of the V$_\text{C}$V$_\text{Si}^0$ in the cubic polytype of SiC, i.e., 3C SiC.

We carried out experiments and \textit{ab initio} calculations on the temperature dependent V$_\text{C}$V$_\text{Si}^-$ $\rightarrow$ V$_\text{C}$V$_\text{Si}^0$ reionization process in 4H SiC. We found that the reionization of divacancies in 4H SiC upon unfocused green repump laser illumination is a complex process in which very efficient traps (most likely nitrogen donors) and the silicon vacancy defects present in the sample are involved. For efficient driving of divacancy qubits in 4H SiC at elevated temperatures, lower excitation energies are preferable. We found that the PL signals of divacancies can be fully recovered at room temperature with typical infared photoexcitation energies without applying any green repump laser for such divacancy configurations which have stable fluorescence up to room temperature. This result strongly indicates that the reionization threshold energies of divacancies decrease with increasing temperature. Our calculations revealed that not the thermal shifts of the band edges nor the charge transition levels of divacancies caused by entropy contributions to the formation enthalpy are responsible for an effective lowering of the reionization threshold energies at elevated temperatures but rather the activation of phonon-assisted photoionization transitions. This result highlights a general conclusion that phonon-assisted photoionization processes should be always considered in understanding the thermally activated photoionization of point defects in semiconductors and insulators.

\begin{acknowledgments}
A.\ Cs.\ acknowledges the support from the \'UNKP-20-4 New National Excellence Program of the Ministry for Innovation and Technology from the source of the National Research, Development and Innovation Fund. A.\ G.\ acknowledges the National Research, Development, and Innovation Office for Hungary Grant No.\ KKP129866 of the National Excellence Program of Quantum-coherent materials project, Grant No.\ 127902 of the EU QuantERA Nanospin project, Grant No.\ 2017-1.2.1-NKP-2017-00001 of the National Quantum Technology Program, and the Quantum Information National Laboratory supported by the Ministry of Innovation and Technology of Hungary, as well as the EU Commission for the H2020 Quantum technology Flagship projects ASTERIQS (Grant No.\ 820394). We acknowledge the EU H2020 project QuanTELCO (Grant No.\ 862721). The support provided by the Swedish Research Council (Grant No.\ VR 2016-05362 for I.\ G.\ I.\ and No.\ VR 2016-04068 for N.\ T.\ S.) and the Knut and Alice Wallenberg Foundation Grant No.\ KAW 2018.0071 for N.\ T.\ S.\ and I.\ G.\ I.\ is acknowledged. We acknowledge the computational sources provided by the Swedish National Infrastructure for Computing (SNIC) at National Computation Centre (NSC) partially funded by the Swedish Research Council through grant agreement No.\ 2018-05973 and the Hungarian Governmental Information Technology Development Agency through the project ``gallium''. We acknowledge that the results of this research have been partially achieved using the DECI resource Eagle HPC based
in Poland at Poznan with support from the PRACE aisbl and resources provided by the Hungarian Governmental Information Technology Development Agency.
\end{acknowledgments}


\begin{thebibliography}{59}%
\makeatletter
\providecommand \@ifxundefined [1]{%
 \@ifx{#1\undefined}
}%
\providecommand \@ifnum [1]{%
 \ifnum #1\expandafter \@firstoftwo
 \else \expandafter \@secondoftwo
 \fi
}%
\providecommand \@ifx [1]{%
 \ifx #1\expandafter \@firstoftwo
 \else \expandafter \@secondoftwo
 \fi
}%
\providecommand \natexlab [1]{#1}%
\providecommand \enquote  [1]{``#1''}%
\providecommand \bibnamefont  [1]{#1}%
\providecommand \bibfnamefont [1]{#1}%
\providecommand \citenamefont [1]{#1}%
\providecommand \href@noop [0]{\@secondoftwo}%
\providecommand \href [0]{\begingroup \@sanitize@url \@href}%
\providecommand \@href[1]{\@@startlink{#1}\@@href}%
\providecommand \@@href[1]{\endgroup#1\@@endlink}%
\providecommand \@sanitize@url [0]{\catcode `\\12\catcode `\$12\catcode
  `\&12\catcode `\#12\catcode `\^12\catcode `\_12\catcode `\%12\relax}%
\providecommand \@@startlink[1]{}%
\providecommand \@@endlink[0]{}%
\providecommand \url  [0]{\begingroup\@sanitize@url \@url }%
\providecommand \@url [1]{\endgroup\@href {#1}{\urlprefix }}%
\providecommand \urlprefix  [0]{URL }%
\providecommand \Eprint [0]{\href }%
\providecommand \doibase [0]{http://dx.doi.org/}%
\providecommand \selectlanguage [0]{\@gobble}%
\providecommand \bibinfo  [0]{\@secondoftwo}%
\providecommand \bibfield  [0]{\@secondoftwo}%
\providecommand \translation [1]{[#1]}%
\providecommand \BibitemOpen [0]{}%
\providecommand \bibitemStop [0]{}%
\providecommand \bibitemNoStop [0]{.\EOS\space}%
\providecommand \EOS [0]{\spacefactor3000\relax}%
\providecommand \BibitemShut  [1]{\csname bibitem#1\endcsname}%
\let\auto@bib@innerbib\@empty
\bibitem [{\citenamefont {Doherty}\ \emph {et~al.}(2013)\citenamefont
  {Doherty}, \citenamefont {Manson}, \citenamefont {Delaney}, \citenamefont
  {Jelezko}, \citenamefont {Wrachtrup},\ and\ \citenamefont
  {Hollenberg}}]{DohertyPhysRep2013}%
  \BibitemOpen
  \bibfield  {author} {\bibinfo {author} {\bibfnamefont {M.~W.}\ \bibnamefont
  {Doherty}}, \bibinfo {author} {\bibfnamefont {N.~B.}\ \bibnamefont {Manson}},
  \bibinfo {author} {\bibfnamefont {P.}~\bibnamefont {Delaney}}, \bibinfo
  {author} {\bibfnamefont {F.}~\bibnamefont {Jelezko}}, \bibinfo {author}
  {\bibfnamefont {J.}~\bibnamefont {Wrachtrup}}, \ and\ \bibinfo {author}
  {\bibfnamefont {L.}~\bibnamefont {Hollenberg}},\ }\href@noop {} {\bibfield
  {journal} {\bibinfo  {journal} {Phys. Rep.}\ }\textbf {\bibinfo {volume}
  {528}},\ \bibinfo {pages} {1} (\bibinfo {year} {2013})}\BibitemShut {NoStop}%
\bibitem [{\citenamefont {Gali}(2019)}]{Gali2019}%
  \BibitemOpen
  \bibfield  {author} {\bibinfo {author} {\bibfnamefont {A.}~\bibnamefont
  {Gali}},\ }\href {\doibase 10.1515/nanoph-2019-0154} {\bibfield  {journal}
  {\bibinfo  {journal} {Nanophotonics}\ }\textbf {\bibinfo {volume} {8}},\
  \bibinfo {pages} {1907} (\bibinfo {year} {2019})}\BibitemShut {NoStop}%
\bibitem [{\citenamefont {Koehl}\ \emph {et~al.}(2011)\citenamefont {Koehl},
  \citenamefont {Buckley}, \citenamefont {Heremans}, \citenamefont {Calusine},\
  and\ \citenamefont {Awschalom}}]{KoehlNature2011}%
  \BibitemOpen
  \bibfield  {author} {\bibinfo {author} {\bibfnamefont {W.~F.}\ \bibnamefont
  {Koehl}}, \bibinfo {author} {\bibfnamefont {B.~B.}\ \bibnamefont {Buckley}},
  \bibinfo {author} {\bibfnamefont {F.~J.}\ \bibnamefont {Heremans}}, \bibinfo
  {author} {\bibfnamefont {G.}~\bibnamefont {Calusine}}, \ and\ \bibinfo
  {author} {\bibfnamefont {D.~D.}\ \bibnamefont {Awschalom}},\ }\href {\doibase
  10.1038/nature10562} {\bibfield  {journal} {\bibinfo  {journal} {Nature}\
  }\textbf {\bibinfo {volume} {479}},\ \bibinfo {pages} {84} (\bibinfo {year}
  {2011})}\BibitemShut {NoStop}%
\bibitem [{\citenamefont {Castelletto}\ \emph {et~al.}(2013)\citenamefont
  {Castelletto}, \citenamefont {Johnson}, \citenamefont {IvĂˇdy}, \citenamefont
  {Stavrias}, \citenamefont {T.}, \citenamefont {Gali},\ and\ \citenamefont
  {Oshima}}]{CastelettoNatMater2013}%
  \BibitemOpen
  \bibfield  {author} {\bibinfo {author} {\bibfnamefont {S.}~\bibnamefont
  {Castelletto}}, \bibinfo {author} {\bibfnamefont {B.~C.}\ \bibnamefont
  {Johnson}}, \bibinfo {author} {\bibfnamefont {V.}~\bibnamefont {IvĂˇdy}},
  \bibinfo {author} {\bibfnamefont {N.}~\bibnamefont {Stavrias}}, \bibinfo
  {author} {\bibfnamefont {U.}~\bibnamefont {T.}}, \bibinfo {author}
  {\bibfnamefont {A.}~\bibnamefont {Gali}}, \ and\ \bibinfo {author}
  {\bibfnamefont {T.}~\bibnamefont {Oshima}},\ }\href {\doibase
  10.1038/nmat3806} {\bibfield  {journal} {\bibinfo  {journal} {Nat. Mater.}\
  }\textbf {\bibinfo {volume} {13}},\ \bibinfo {pages} {151} (\bibinfo {year}
  {2013})}\BibitemShut {NoStop}%
\bibitem [{\citenamefont {Christle}\ \emph {et~al.}(2015)\citenamefont
  {Christle}, \citenamefont {Falk}, \citenamefont {P.}, \citenamefont {Klimov},
  \citenamefont {Ul~Hassan}, \citenamefont {Son}, \citenamefont {E.},
  \citenamefont {T.},\ and\ \citenamefont {Awschalom}}]{ChristleNatMater2015}%
  \BibitemOpen
  \bibfield  {author} {\bibinfo {author} {\bibfnamefont {D.~J.}\ \bibnamefont
  {Christle}}, \bibinfo {author} {\bibfnamefont {A.~L.}\ \bibnamefont {Falk}},
  \bibinfo {author} {\bibfnamefont {A.}~\bibnamefont {P.}}, \bibinfo {author}
  {\bibfnamefont {P.~V.}\ \bibnamefont {Klimov}}, \bibinfo {author}
  {\bibfnamefont {J.}~\bibnamefont {Ul~Hassan}}, \bibinfo {author}
  {\bibfnamefont {N.~T.}\ \bibnamefont {Son}}, \bibinfo {author} {\bibfnamefont
  {J.}~\bibnamefont {E.}}, \bibinfo {author} {\bibfnamefont {O.}~\bibnamefont
  {T.}}, \ and\ \bibinfo {author} {\bibfnamefont {D.~D.}\ \bibnamefont
  {Awschalom}},\ }\href {\doibase 10.1038/nmat4145} {\bibfield  {journal}
  {\bibinfo  {journal} {Nat. Mater.}\ }\textbf {\bibinfo {volume} {14}},\
  \bibinfo {pages} {164} (\bibinfo {year} {2015})}\BibitemShut {NoStop}%
\bibitem [{\citenamefont {Widmann}\ \emph {et~al.}(2015)\citenamefont
  {Widmann}, \citenamefont {Lee}, \citenamefont {Rendler}, \citenamefont {Son},
  \citenamefont {Fedder}, \citenamefont {Paik}, \citenamefont {Yang},
  \citenamefont {Zhao}, \citenamefont {Yang}, \citenamefont {Booker},
  \citenamefont {Denisenko}, \citenamefont {Jamali}, \citenamefont
  {Momenzadeh}, \citenamefont {Gerhardt}, \citenamefont {Ohshima},
  \citenamefont {Gali}, \citenamefont {Janz\'en},\ and\ \citenamefont
  {Wrachtrup}}]{Widmann2015NatMat}%
  \BibitemOpen
  \bibfield  {author} {\bibinfo {author} {\bibfnamefont {M.}~\bibnamefont
  {Widmann}}, \bibinfo {author} {\bibfnamefont {S.-Y.}\ \bibnamefont {Lee}},
  \bibinfo {author} {\bibfnamefont {T.}~\bibnamefont {Rendler}}, \bibinfo
  {author} {\bibfnamefont {N.~T.}\ \bibnamefont {Son}}, \bibinfo {author}
  {\bibfnamefont {H.}~\bibnamefont {Fedder}}, \bibinfo {author} {\bibfnamefont
  {S.}~\bibnamefont {Paik}}, \bibinfo {author} {\bibfnamefont {L.-P.}\
  \bibnamefont {Yang}}, \bibinfo {author} {\bibfnamefont {N.}~\bibnamefont
  {Zhao}}, \bibinfo {author} {\bibfnamefont {S.}~\bibnamefont {Yang}}, \bibinfo
  {author} {\bibfnamefont {I.}~\bibnamefont {Booker}}, \bibinfo {author}
  {\bibfnamefont {A.}~\bibnamefont {Denisenko}}, \bibinfo {author}
  {\bibfnamefont {M.}~\bibnamefont {Jamali}}, \bibinfo {author} {\bibfnamefont
  {S.~A.}\ \bibnamefont {Momenzadeh}}, \bibinfo {author} {\bibfnamefont
  {I.}~\bibnamefont {Gerhardt}}, \bibinfo {author} {\bibfnamefont
  {T.}~\bibnamefont {Ohshima}}, \bibinfo {author} {\bibfnamefont
  {A.}~\bibnamefont {Gali}}, \bibinfo {author} {\bibfnamefont {E.}~\bibnamefont
  {Janz\'en}}, \ and\ \bibinfo {author} {\bibfnamefont {J.}~\bibnamefont
  {Wrachtrup}},\ }\href {\doibase 10.1038/nmat4145} {\bibfield  {journal}
  {\bibinfo  {journal} {Nature Materials}\ }\textbf {\bibinfo {volume} {14}},\
  \bibinfo {pages} {164} (\bibinfo {year} {2015})}\BibitemShut {NoStop}%
\bibitem [{\citenamefont {Wolfowicz}\ \emph {et~al.}(2020)\citenamefont
  {Wolfowicz}, \citenamefont {Anderson}, \citenamefont {Diler}, \citenamefont
  {Poluektov}, \citenamefont {Heremans},\ and\ \citenamefont
  {Awschalom}}]{Wolfowicz2020}%
  \BibitemOpen
  \bibfield  {author} {\bibinfo {author} {\bibfnamefont {G.}~\bibnamefont
  {Wolfowicz}}, \bibinfo {author} {\bibfnamefont {C.~P.}\ \bibnamefont
  {Anderson}}, \bibinfo {author} {\bibfnamefont {B.}~\bibnamefont {Diler}},
  \bibinfo {author} {\bibfnamefont {O.~G.}\ \bibnamefont {Poluektov}}, \bibinfo
  {author} {\bibfnamefont {F.~J.}\ \bibnamefont {Heremans}}, \ and\ \bibinfo
  {author} {\bibfnamefont {D.~D.}\ \bibnamefont {Awschalom}},\ }\href {\doibase
  10.1126/sciadv.aaz1192} {\bibfield  {journal} {\bibinfo  {journal} {Science
  Advances}\ }\textbf {\bibinfo {volume} {6}},\ \bibinfo {pages} {eaaz1192}
  (\bibinfo {year} {2020})},\ \bibinfo {note} {publisher: American Association
  for the Advancement of Science Section: Research Article}\BibitemShut
  {NoStop}%
\bibitem [{\citenamefont {Lohrmann}\ \emph {et~al.}(2015)\citenamefont
  {Lohrmann}, \citenamefont {Iwamoto}, \citenamefont {Bodrog}, \citenamefont
  {Castelletto}, \citenamefont {Ohshima}, \citenamefont {Karle}, \citenamefont
  {Gali}, \citenamefont {S.}, \citenamefont {McCallum},\ and\ \citenamefont
  {Johnson}}]{LohrmannNatComm2015}%
  \BibitemOpen
  \bibfield  {author} {\bibinfo {author} {\bibfnamefont {A.}~\bibnamefont
  {Lohrmann}}, \bibinfo {author} {\bibfnamefont {N.}~\bibnamefont {Iwamoto}},
  \bibinfo {author} {\bibfnamefont {Z.}~\bibnamefont {Bodrog}}, \bibinfo
  {author} {\bibfnamefont {S.}~\bibnamefont {Castelletto}}, \bibinfo {author}
  {\bibfnamefont {T.}~\bibnamefont {Ohshima}}, \bibinfo {author} {\bibfnamefont
  {T.~J.}\ \bibnamefont {Karle}}, \bibinfo {author} {\bibfnamefont
  {A.}~\bibnamefont {Gali}}, \bibinfo {author} {\bibfnamefont {P.}~\bibnamefont
  {S.}}, \bibinfo {author} {\bibfnamefont {J.~C.}\ \bibnamefont {McCallum}}, \
  and\ \bibinfo {author} {\bibfnamefont {B.~C.}\ \bibnamefont {Johnson}},\
  }\href {\doibase 10.1038/ncomms8783} {\bibfield  {journal} {\bibinfo
  {journal} {Nat. Comm.}\ }\textbf {\bibinfo {volume} {6}},\ \bibinfo {pages}
  {7783} (\bibinfo {year} {2015})}\BibitemShut {NoStop}%
\bibitem [{\citenamefont {Radulaski}\ \emph {et~al.}(2017)\citenamefont
  {Radulaski}, \citenamefont {Widmann}, \citenamefont {Niethammer},
  \citenamefont {Zhang}, \citenamefont {Lee}, \citenamefont {Rendler},
  \citenamefont {Lagoudakis}, \citenamefont {Son}, \citenamefont {Janz\'en},
  \citenamefont {Ohshima}, \citenamefont {Wrachtrup},\ and\ \citenamefont
  {Vu\v{c}kovi\.{c}}}]{RadulaskiNanoLett2017}%
  \BibitemOpen
  \bibfield  {author} {\bibinfo {author} {\bibfnamefont {M.}~\bibnamefont
  {Radulaski}}, \bibinfo {author} {\bibfnamefont {M.}~\bibnamefont {Widmann}},
  \bibinfo {author} {\bibfnamefont {M.}~\bibnamefont {Niethammer}}, \bibinfo
  {author} {\bibfnamefont {J.~L.}\ \bibnamefont {Zhang}}, \bibinfo {author}
  {\bibfnamefont {S.-Y.}\ \bibnamefont {Lee}}, \bibinfo {author} {\bibfnamefont
  {T.}~\bibnamefont {Rendler}}, \bibinfo {author} {\bibfnamefont {K.~G.}\
  \bibnamefont {Lagoudakis}}, \bibinfo {author} {\bibfnamefont {N.~T.}\
  \bibnamefont {Son}}, \bibinfo {author} {\bibfnamefont {E.}~\bibnamefont
  {Janz\'en}}, \bibinfo {author} {\bibfnamefont {T.}~\bibnamefont {Ohshima}},
  \bibinfo {author} {\bibfnamefont {J.}~\bibnamefont {Wrachtrup}}, \ and\
  \bibinfo {author} {\bibfnamefont {J.}~\bibnamefont {Vu\v{c}kovi\.{c}}},\
  }\href {\doibase 10.1021/acs.nanolett.6b05102} {\bibfield  {journal}
  {\bibinfo  {journal} {Nano Letters}\ }\textbf {\bibinfo {volume} {17}},\
  \bibinfo {pages} {1782} (\bibinfo {year} {2017})},\ \bibinfo {note} {pMID:
  28225630},\ \Eprint
  {http://arxiv.org/abs/https://doi.org/10.1021/acs.nanolett.6b05102}
  {https://doi.org/10.1021/acs.nanolett.6b05102} \BibitemShut {NoStop}%
\bibitem [{\citenamefont {Spindlberger}\ \emph {et~al.}(2019)\citenamefont
  {Spindlberger}, \citenamefont {Cs\'or\'e}, \citenamefont {Thiering},
  \citenamefont {Putz}, \citenamefont {Karhu}, \citenamefont {Hassan},
  \citenamefont {Son}, \citenamefont {Fromherz}, \citenamefont {Gali},\ and\
  \citenamefont {Trupke}}]{SpindlbergerPRA2019}%
  \BibitemOpen
  \bibfield  {author} {\bibinfo {author} {\bibfnamefont {L.}~\bibnamefont
  {Spindlberger}}, \bibinfo {author} {\bibfnamefont {A.}~\bibnamefont
  {Cs\'or\'e}}, \bibinfo {author} {\bibfnamefont {G.}~\bibnamefont {Thiering}},
  \bibinfo {author} {\bibfnamefont {S.}~\bibnamefont {Putz}}, \bibinfo {author}
  {\bibfnamefont {R.}~\bibnamefont {Karhu}}, \bibinfo {author} {\bibfnamefont
  {J.}~\bibnamefont {Hassan}}, \bibinfo {author} {\bibfnamefont
  {N.}~\bibnamefont {Son}}, \bibinfo {author} {\bibfnamefont {T.}~\bibnamefont
  {Fromherz}}, \bibinfo {author} {\bibfnamefont {A.}~\bibnamefont {Gali}}, \
  and\ \bibinfo {author} {\bibfnamefont {M.}~\bibnamefont {Trupke}},\ }\href
  {\doibase 10.1103/PhysRevApplied.12.014015} {\bibfield  {journal} {\bibinfo
  {journal} {Phys. Rev. Applied}\ }\textbf {\bibinfo {volume} {12}},\ \bibinfo
  {pages} {014015} (\bibinfo {year} {2019})}\BibitemShut {NoStop}%
\bibitem [{\citenamefont {Nagy}\ \emph {et~al.}(2019)\citenamefont {Nagy},
  \citenamefont {Niethammer}, \citenamefont {Widmann}, \citenamefont {Chen},
  \citenamefont {Udvarhelyi}, \citenamefont {Bonato}, \citenamefont {Hassan},
  \citenamefont {Karhu}, \citenamefont {Ivanov}, \citenamefont {Son},
  \citenamefont {Maze}, \citenamefont {Ohshima}, \citenamefont {Soykal},
  \citenamefont {Gali}, \citenamefont {Lee}, \citenamefont {Kaiser},\ and\
  \citenamefont {Wrachtrup}}]{Nagy2019}%
  \BibitemOpen
  \bibfield  {author} {\bibinfo {author} {\bibfnamefont {R.}~\bibnamefont
  {Nagy}}, \bibinfo {author} {\bibfnamefont {M.}~\bibnamefont {Niethammer}},
  \bibinfo {author} {\bibfnamefont {M.}~\bibnamefont {Widmann}}, \bibinfo
  {author} {\bibfnamefont {Y.-C.}\ \bibnamefont {Chen}}, \bibinfo {author}
  {\bibfnamefont {P.}~\bibnamefont {Udvarhelyi}}, \bibinfo {author}
  {\bibfnamefont {C.}~\bibnamefont {Bonato}}, \bibinfo {author} {\bibfnamefont
  {J.~U.}\ \bibnamefont {Hassan}}, \bibinfo {author} {\bibfnamefont
  {R.}~\bibnamefont {Karhu}}, \bibinfo {author} {\bibfnamefont {I.~G.}\
  \bibnamefont {Ivanov}}, \bibinfo {author} {\bibfnamefont {N.~T.}\
  \bibnamefont {Son}}, \bibinfo {author} {\bibfnamefont {J.~R.}\ \bibnamefont
  {Maze}}, \bibinfo {author} {\bibfnamefont {T.}~\bibnamefont {Ohshima}},
  \bibinfo {author} {\bibfnamefont {O.~O.}\ \bibnamefont {Soykal}}, \bibinfo
  {author} {\bibfnamefont {A.}~\bibnamefont {Gali}}, \bibinfo {author}
  {\bibfnamefont {S.-Y.}\ \bibnamefont {Lee}}, \bibinfo {author} {\bibfnamefont
  {F.}~\bibnamefont {Kaiser}}, \ and\ \bibinfo {author} {\bibfnamefont
  {J.}~\bibnamefont {Wrachtrup}},\ }\href {\doibase 10.1038/s41467-019-09873-9}
  {\bibfield  {journal} {\bibinfo  {journal} {Nature Communications}\ }\textbf
  {\bibinfo {volume} {10}},\ \bibinfo {pages} {1954} (\bibinfo {year}
  {2019})}\BibitemShut {NoStop}%
\bibitem [{\citenamefont {Babin}\ \emph {et~al.}(2022)\citenamefont {Babin},
  \citenamefont {St\"ohr}, \citenamefont {Morioka}, \citenamefont {Linkewitz},
  \citenamefont {Steidl}, \citenamefont {W\"ornle}, \citenamefont {Liu},
  \citenamefont {Hesselmeier}, \citenamefont {Vorobyov}, \citenamefont
  {Denisenko}, \citenamefont {Hentschel}, \citenamefont {Gobert}, \citenamefont
  {Berwian}, \citenamefont {Astakhov}, \citenamefont {Knolle}, \citenamefont
  {Majety}, \citenamefont {Saha}, \citenamefont {Radulaski}, \citenamefont
  {Son}, \citenamefont {Ul-Hassan}, \citenamefont {Kaiser},\ and\ \citenamefont
  {Wrachtrup}}]{Babin2022}%
  \BibitemOpen
  \bibfield  {author} {\bibinfo {author} {\bibfnamefont {C.}~\bibnamefont
  {Babin}}, \bibinfo {author} {\bibfnamefont {R.}~\bibnamefont {St\"ohr}},
  \bibinfo {author} {\bibfnamefont {N.}~\bibnamefont {Morioka}}, \bibinfo
  {author} {\bibfnamefont {T.}~\bibnamefont {Linkewitz}}, \bibinfo {author}
  {\bibfnamefont {T.}~\bibnamefont {Steidl}}, \bibinfo {author} {\bibfnamefont
  {R.}~\bibnamefont {W\"ornle}}, \bibinfo {author} {\bibfnamefont
  {D.}~\bibnamefont {Liu}}, \bibinfo {author} {\bibfnamefont {E.}~\bibnamefont
  {Hesselmeier}}, \bibinfo {author} {\bibfnamefont {V.}~\bibnamefont
  {Vorobyov}}, \bibinfo {author} {\bibfnamefont {A.}~\bibnamefont {Denisenko}},
  \bibinfo {author} {\bibfnamefont {M.}~\bibnamefont {Hentschel}}, \bibinfo
  {author} {\bibfnamefont {C.}~\bibnamefont {Gobert}}, \bibinfo {author}
  {\bibfnamefont {P.}~\bibnamefont {Berwian}}, \bibinfo {author} {\bibfnamefont
  {G.~V.}\ \bibnamefont {Astakhov}}, \bibinfo {author} {\bibfnamefont
  {W.}~\bibnamefont {Knolle}}, \bibinfo {author} {\bibfnamefont
  {S.}~\bibnamefont {Majety}}, \bibinfo {author} {\bibfnamefont
  {P.}~\bibnamefont {Saha}}, \bibinfo {author} {\bibfnamefont {M.}~\bibnamefont
  {Radulaski}}, \bibinfo {author} {\bibfnamefont {N.~T.}\ \bibnamefont {Son}},
  \bibinfo {author} {\bibfnamefont {J.}~\bibnamefont {Ul-Hassan}}, \bibinfo
  {author} {\bibfnamefont {F.}~\bibnamefont {Kaiser}}, \ and\ \bibinfo {author}
  {\bibfnamefont {J.}~\bibnamefont {Wrachtrup}},\ }\href {\doibase
  10.1038/s41563-021-01148-3} {\bibfield  {journal} {\bibinfo  {journal}
  {Nature Materials}\ }\textbf {\bibinfo {volume} {21}},\ \bibinfo {pages} {67}
  (\bibinfo {year} {2022})},\ \bibinfo {note} {number: 1 Publisher: Nature
  Publishing Group}\BibitemShut {NoStop}%
\bibitem [{\citenamefont {Lee}\ \emph {et~al.}(2015)\citenamefont {Lee},
  \citenamefont {Niethammer},\ and\ \citenamefont
  {Wrachtrup}}]{SangYunPRB2015}%
  \BibitemOpen
  \bibfield  {author} {\bibinfo {author} {\bibfnamefont {S.-Y.}\ \bibnamefont
  {Lee}}, \bibinfo {author} {\bibfnamefont {M.}~\bibnamefont {Niethammer}}, \
  and\ \bibinfo {author} {\bibfnamefont {J.}~\bibnamefont {Wrachtrup}},\ }\href
  {\doibase 10.1103/PhysRevB.92.115201} {\bibfield  {journal} {\bibinfo
  {journal} {Phys. Rev. B}\ }\textbf {\bibinfo {volume} {92}},\ \bibinfo
  {pages} {115201} (\bibinfo {year} {2015})}\BibitemShut {NoStop}%
\bibitem [{\citenamefont {Kraus}\ \emph {et~al.}(2014)\citenamefont {Kraus},
  \citenamefont {Soltamov}, \citenamefont {Fuchs}, \citenamefont {Simin},
  \citenamefont {Sperlich}, \citenamefont {Baranov}, \citenamefont {Astakhov},\
  and\ \citenamefont {Dyakonov}}]{KrausSciRep2014}%
  \BibitemOpen
  \bibfield  {author} {\bibinfo {author} {\bibfnamefont {H.}~\bibnamefont
  {Kraus}}, \bibinfo {author} {\bibfnamefont {V.~A.}\ \bibnamefont {Soltamov}},
  \bibinfo {author} {\bibfnamefont {F.}~\bibnamefont {Fuchs}}, \bibinfo
  {author} {\bibfnamefont {D.}~\bibnamefont {Simin}}, \bibinfo {author}
  {\bibfnamefont {A.}~\bibnamefont {Sperlich}}, \bibinfo {author}
  {\bibfnamefont {P.~G.}\ \bibnamefont {Baranov}}, \bibinfo {author}
  {\bibfnamefont {G.~V.}\ \bibnamefont {Astakhov}}, \ and\ \bibinfo {author}
  {\bibfnamefont {V.}~\bibnamefont {Dyakonov}},\ }\href {\doibase
  10.1038/srep05303} {\bibfield  {journal} {\bibinfo  {journal} {Scientific
  Reports}\ }\textbf {\bibinfo {volume} {4}} (\bibinfo {year} {2014}),\
  10.1038/srep05303}\BibitemShut {NoStop}%
\bibitem [{\citenamefont {Simin}\ \emph {et~al.}(2016)\citenamefont {Simin},
  \citenamefont {Soltamov}, \citenamefont {Poshakinskiy}, \citenamefont
  {Anisimov}, \citenamefont {Babunts}, \citenamefont {Tolmachev}, \citenamefont
  {Mokhov}, \citenamefont {Trupke}, \citenamefont {Tarasenko}, \citenamefont
  {Sperlich}, \citenamefont {Baranov}, \citenamefont {Dyakonov},\ and\
  \citenamefont {Astakhov}}]{Simin2016PRX}%
  \BibitemOpen
  \bibfield  {author} {\bibinfo {author} {\bibfnamefont {D.}~\bibnamefont
  {Simin}}, \bibinfo {author} {\bibfnamefont {V.~A.}\ \bibnamefont {Soltamov}},
  \bibinfo {author} {\bibfnamefont {A.~V.}\ \bibnamefont {Poshakinskiy}},
  \bibinfo {author} {\bibfnamefont {A.~N.}\ \bibnamefont {Anisimov}}, \bibinfo
  {author} {\bibfnamefont {R.~A.}\ \bibnamefont {Babunts}}, \bibinfo {author}
  {\bibfnamefont {D.~O.}\ \bibnamefont {Tolmachev}}, \bibinfo {author}
  {\bibfnamefont {E.~N.}\ \bibnamefont {Mokhov}}, \bibinfo {author}
  {\bibfnamefont {M.}~\bibnamefont {Trupke}}, \bibinfo {author} {\bibfnamefont
  {S.~A.}\ \bibnamefont {Tarasenko}}, \bibinfo {author} {\bibfnamefont
  {A.}~\bibnamefont {Sperlich}}, \bibinfo {author} {\bibfnamefont {P.~G.}\
  \bibnamefont {Baranov}}, \bibinfo {author} {\bibfnamefont {V.}~\bibnamefont
  {Dyakonov}}, \ and\ \bibinfo {author} {\bibfnamefont {G.~V.}\ \bibnamefont
  {Astakhov}},\ }\href {\doibase 10.1103/PhysRevX.6.031014} {\bibfield
  {journal} {\bibinfo  {journal} {Physical Review X}\ }\textbf {\bibinfo
  {volume} {6}},\ \bibinfo {pages} {031014} (\bibinfo {year}
  {2016})}\BibitemShut {NoStop}%
\bibitem [{\citenamefont {Simin}\ \emph {et~al.}(2015)\citenamefont {Simin},
  \citenamefont {Fuchs}, \citenamefont {Kraus}, \citenamefont {Sperlich},
  \citenamefont {Baranov}, \citenamefont {Astakhov},\ and\ \citenamefont
  {Dyakonov}}]{SiminPRApplied2015}%
  \BibitemOpen
  \bibfield  {author} {\bibinfo {author} {\bibfnamefont {D.}~\bibnamefont
  {Simin}}, \bibinfo {author} {\bibfnamefont {F.}~\bibnamefont {Fuchs}},
  \bibinfo {author} {\bibfnamefont {H.}~\bibnamefont {Kraus}}, \bibinfo
  {author} {\bibfnamefont {A.}~\bibnamefont {Sperlich}}, \bibinfo {author}
  {\bibfnamefont {P.~G.}\ \bibnamefont {Baranov}}, \bibinfo {author}
  {\bibfnamefont {G.~V.}\ \bibnamefont {Astakhov}}, \ and\ \bibinfo {author}
  {\bibfnamefont {V.}~\bibnamefont {Dyakonov}},\ }\href {\doibase
  10.1103/PhysRevApplied.4.014009} {\bibfield  {journal} {\bibinfo  {journal}
  {Physical Review Applied}\ }\textbf {\bibinfo {volume} {4}},\ \bibinfo
  {pages} {014009} (\bibinfo {year} {2015})}\BibitemShut {NoStop}%
\bibitem [{\citenamefont {Niethammer}\ \emph {et~al.}(2016)\citenamefont
  {Niethammer}, \citenamefont {Widmann}, \citenamefont {Lee}, \citenamefont
  {Stenberg}, \citenamefont {Kordina}, \citenamefont {Ohshima}, \citenamefont
  {Son}, \citenamefont {Janz\'en},\ and\ \citenamefont
  {Wrachtrup}}]{NiethammerPRApplied2016}%
  \BibitemOpen
  \bibfield  {author} {\bibinfo {author} {\bibfnamefont {M.}~\bibnamefont
  {Niethammer}}, \bibinfo {author} {\bibfnamefont {M.}~\bibnamefont {Widmann}},
  \bibinfo {author} {\bibfnamefont {S.-Y.}\ \bibnamefont {Lee}}, \bibinfo
  {author} {\bibfnamefont {P.}~\bibnamefont {Stenberg}}, \bibinfo {author}
  {\bibfnamefont {O.}~\bibnamefont {Kordina}}, \bibinfo {author} {\bibfnamefont
  {T.}~\bibnamefont {Ohshima}}, \bibinfo {author} {\bibfnamefont {N.~T.}\
  \bibnamefont {Son}}, \bibinfo {author} {\bibfnamefont {E.}~\bibnamefont
  {Janz\'en}}, \ and\ \bibinfo {author} {\bibfnamefont {J.}~\bibnamefont
  {Wrachtrup}},\ }\href {\doibase 10.1103/PhysRevApplied.6.034001} {\bibfield
  {journal} {\bibinfo  {journal} {Physical Review Applied}\ }\textbf {\bibinfo
  {volume} {6}},\ \bibinfo {pages} {034001} (\bibinfo {year}
  {2016})}\BibitemShut {NoStop}%
\bibitem [{\citenamefont {Cochrane}\ \emph {et~al.}(2016)\citenamefont
  {Cochrane}, \citenamefont {Blacksberg}, \citenamefont {Anders},\ and\
  \citenamefont {Lenahan}}]{CochraneSciRep2016}%
  \BibitemOpen
  \bibfield  {author} {\bibinfo {author} {\bibfnamefont {C.~J.}\ \bibnamefont
  {Cochrane}}, \bibinfo {author} {\bibfnamefont {J.}~\bibnamefont
  {Blacksberg}}, \bibinfo {author} {\bibfnamefont {M.~A.}\ \bibnamefont
  {Anders}}, \ and\ \bibinfo {author} {\bibfnamefont {P.~M.}\ \bibnamefont
  {Lenahan}},\ }\href {\doibase 10.1038/srep37077} {\bibfield  {journal}
  {\bibinfo  {journal} {Scientific Reports}\ }\textbf {\bibinfo {volume} {6}},\
  \bibinfo {pages} {37077} (\bibinfo {year} {2016})}\BibitemShut {NoStop}%
\bibitem [{\citenamefont {Anisimov}\ \emph {et~al.}(2016)\citenamefont
  {Anisimov}, \citenamefont {Simin}, \citenamefont {Soltamov}, \citenamefont
  {Lebedev}, \citenamefont {Baranov}, \citenamefont {Astakhov},\ and\
  \citenamefont {Dyakonov}}]{AnisimovSciRep2016}%
  \BibitemOpen
  \bibfield  {author} {\bibinfo {author} {\bibfnamefont {A.~N.}\ \bibnamefont
  {Anisimov}}, \bibinfo {author} {\bibfnamefont {D.}~\bibnamefont {Simin}},
  \bibinfo {author} {\bibfnamefont {V.~A.}\ \bibnamefont {Soltamov}}, \bibinfo
  {author} {\bibfnamefont {S.~P.}\ \bibnamefont {Lebedev}}, \bibinfo {author}
  {\bibfnamefont {P.~G.}\ \bibnamefont {Baranov}}, \bibinfo {author}
  {\bibfnamefont {G.~V.}\ \bibnamefont {Astakhov}}, \ and\ \bibinfo {author}
  {\bibfnamefont {V.}~\bibnamefont {Dyakonov}},\ }\href
  {http://dx.doi.org/10.1038/srep33301} {\bibfield  {journal} {\bibinfo
  {journal} {Scientific Reports}\ }\textbf {\bibinfo {volume} {6}},\ \bibinfo
  {pages} {33301} (\bibinfo {year} {2016})}\BibitemShut {NoStop}%
\bibitem [{\citenamefont {Gali}(2011)}]{Gali2011}%
  \BibitemOpen
  \bibfield  {author} {\bibinfo {author} {\bibfnamefont {A.}~\bibnamefont
  {Gali}},\ }\href {\doibase 10.1002/pssb.201046254} {\bibfield  {journal}
  {\bibinfo  {journal} {physica status solidi (b)}\ }\textbf {\bibinfo {volume}
  {248}},\ \bibinfo {pages} {1337} (\bibinfo {year} {2011})}\BibitemShut
  {NoStop}%
\bibitem [{\citenamefont {Magnusson}\ and\ \citenamefont
  {Janz\'en}(2005)}]{Magnusson2005}%
  \BibitemOpen
  \bibfield  {author} {\bibinfo {author} {\bibfnamefont {B.}~\bibnamefont
  {Magnusson}}\ and\ \bibinfo {author} {\bibfnamefont {E.}~\bibnamefont
  {Janz\'en}},\ }\href {\doibase 10.4028/www.scientific.net/MSF.483-485.341}
  {\bibfield  {journal} {\bibinfo  {journal} {Materials Science Forum}\
  }\textbf {\bibinfo {volume} {483-485}},\ \bibinfo {pages} {341} (\bibinfo
  {year} {2005})},\ \bibinfo {note} {conference Name: Silicon Carbide and
  Related Materials 2004 ISBN: 9780878499632 Publisher: Trans Tech Publications
  Ltd}\BibitemShut {NoStop}%
\bibitem [{\citenamefont {Baranov}\ \emph {et~al.}(2005)\citenamefont
  {Baranov}, \citenamefont {Il'in}, \citenamefont {Mokhov}, \citenamefont
  {Muzafarova}, \citenamefont {Orlinskii},\ and\ \citenamefont
  {Schmidt}}]{BaranovJETP2005}%
  \BibitemOpen
  \bibfield  {author} {\bibinfo {author} {\bibfnamefont {P.~G.}\ \bibnamefont
  {Baranov}}, \bibinfo {author} {\bibfnamefont {I.~V.}\ \bibnamefont {Il'in}},
  \bibinfo {author} {\bibfnamefont {E.~N.}\ \bibnamefont {Mokhov}}, \bibinfo
  {author} {\bibfnamefont {M.~V.}\ \bibnamefont {Muzafarova}}, \bibinfo
  {author} {\bibfnamefont {S.~B.}\ \bibnamefont {Orlinskii}}, \ and\ \bibinfo
  {author} {\bibfnamefont {J.}~\bibnamefont {Schmidt}},\ }\href {\doibase
  10.1134/1.2142873} {\bibfield  {journal} {\bibinfo  {journal} {Jounal of
  Experimental and Theoretical Physics Letters}\ }\textbf {\bibinfo {volume}
  {82}},\ \bibinfo {pages} {441} (\bibinfo {year} {2005})}\BibitemShut
  {NoStop}%
\bibitem [{\citenamefont {Son}\ \emph {et~al.}(2006)\citenamefont {Son},
  \citenamefont {Carlsson}, \citenamefont {ul~Hassan}, \citenamefont
  {Janz\'en}, \citenamefont {Umeda}, \citenamefont {Isoya}, \citenamefont
  {Gali}, \citenamefont {Bockstedte}, \citenamefont {Morishita}, \citenamefont
  {Ohshima},\ and\ \citenamefont {Itoh}}]{SonPRL2006}%
  \BibitemOpen
  \bibfield  {author} {\bibinfo {author} {\bibfnamefont {N.~T.}\ \bibnamefont
  {Son}}, \bibinfo {author} {\bibfnamefont {P.}~\bibnamefont {Carlsson}},
  \bibinfo {author} {\bibfnamefont {J.}~\bibnamefont {ul~Hassan}}, \bibinfo
  {author} {\bibfnamefont {E.}~\bibnamefont {Janz\'en}}, \bibinfo {author}
  {\bibfnamefont {T.}~\bibnamefont {Umeda}}, \bibinfo {author} {\bibfnamefont
  {J.}~\bibnamefont {Isoya}}, \bibinfo {author} {\bibfnamefont
  {A.}~\bibnamefont {Gali}}, \bibinfo {author} {\bibfnamefont {M.}~\bibnamefont
  {Bockstedte}}, \bibinfo {author} {\bibfnamefont {N.}~\bibnamefont
  {Morishita}}, \bibinfo {author} {\bibfnamefont {T.}~\bibnamefont {Ohshima}},
  \ and\ \bibinfo {author} {\bibfnamefont {H.}~\bibnamefont {Itoh}},\ }\href
  {\doibase 10.1103/PhysRevLett.96.055501} {\bibfield  {journal} {\bibinfo
  {journal} {Phys. Rev. Lett.}\ }\textbf {\bibinfo {volume} {96}},\ \bibinfo
  {pages} {055501} (\bibinfo {year} {2006})}\BibitemShut {NoStop}%
\bibitem [{\citenamefont {Somogyi}\ and\ \citenamefont
  {Gali}(2014)}]{SomogyiIOP2014}%
  \BibitemOpen
  \bibfield  {author} {\bibinfo {author} {\bibfnamefont {B.}~\bibnamefont
  {Somogyi}}\ and\ \bibinfo {author} {\bibfnamefont {A.}~\bibnamefont {Gali}},\
  }\href {\doibase 10.1088/0953-8984/26/14/143202} {\bibfield  {journal}
  {\bibinfo  {journal} {Journal of Physics: Condensed Matter}\ }\textbf
  {\bibinfo {volume} {26}},\ \bibinfo {pages} {143202} (\bibinfo {year}
  {2014})}\BibitemShut {NoStop}%
\bibitem [{\citenamefont {Somogyi}\ \emph {et~al.}(2012)\citenamefont
  {Somogyi}, \citenamefont {Zolyomi},\ and\ \citenamefont
  {Gali}}]{SomogyiNanoscale2012}%
  \BibitemOpen
  \bibfield  {author} {\bibinfo {author} {\bibfnamefont {B.}~\bibnamefont
  {Somogyi}}, \bibinfo {author} {\bibfnamefont {V.}~\bibnamefont {Zolyomi}}, \
  and\ \bibinfo {author} {\bibfnamefont {A.}~\bibnamefont {Gali}},\ }\href
  {\doibase 10.1039/C2NR32442C} {\bibfield  {journal} {\bibinfo  {journal}
  {Nanoscale}\ }\textbf {\bibinfo {volume} {4}},\ \bibinfo {pages} {7720}
  (\bibinfo {year} {2012})}\BibitemShut {NoStop}%
\bibitem [{\citenamefont {Beke}\ \emph {et~al.}(2020)\citenamefont {Beke},
  \citenamefont {Valenta}, \citenamefont {KĂˇrolyhĂˇzy}, \citenamefont {Lenk},
  \citenamefont {CzigĂˇny}, \citenamefont {MĂˇrkus}, \citenamefont {KamarĂˇs},
  \citenamefont {Simon},\ and\ \citenamefont {Gali}}]{BekeJPCL2020}%
  \BibitemOpen
  \bibfield  {author} {\bibinfo {author} {\bibfnamefont {D.}~\bibnamefont
  {Beke}}, \bibinfo {author} {\bibfnamefont {J.}~\bibnamefont {Valenta}},
  \bibinfo {author} {\bibfnamefont {G.}~\bibnamefont {KĂˇrolyhĂˇzy}}, \bibinfo
  {author} {\bibfnamefont {S.}~\bibnamefont {Lenk}}, \bibinfo {author}
  {\bibfnamefont {Z.}~\bibnamefont {CzigĂˇny}}, \bibinfo {author}
  {\bibfnamefont {B.~G.}\ \bibnamefont {MĂˇrkus}}, \bibinfo {author}
  {\bibfnamefont {K.}~\bibnamefont {KamarĂˇs}}, \bibinfo {author}
  {\bibfnamefont {F.}~\bibnamefont {Simon}}, \ and\ \bibinfo {author}
  {\bibfnamefont {A.}~\bibnamefont {Gali}},\ }\href {\doibase
  10.1021/acs.jpclett.0c00052} {\bibfield  {journal} {\bibinfo  {journal} {The
  Journal of Physical Chemistry Letters}\ }\textbf {\bibinfo {volume} {11}},\
  \bibinfo {pages} {1675} (\bibinfo {year} {2020})},\ \bibinfo {note}
  {publisher: American Chemical Society}\BibitemShut {NoStop}%
\bibitem [{\citenamefont {Christle}\ \emph {et~al.}(2017)\citenamefont
  {Christle}, \citenamefont {Klimov}, \citenamefont {de~las Casas},
  \citenamefont {Sz\'asz}, \citenamefont {Iv\'ady}, \citenamefont
  {Jokubavicius}, \citenamefont {Ul~Hassan}, \citenamefont {Syv\"aj\"arvi},
  \citenamefont {Koehl}, \citenamefont {Ohshima}, \citenamefont {Son},
  \citenamefont {Janz\'en}, \citenamefont {Gali},\ and\ \citenamefont
  {Awschalom}}]{ChristlePRX2017}%
  \BibitemOpen
  \bibfield  {author} {\bibinfo {author} {\bibfnamefont {D.~J.}\ \bibnamefont
  {Christle}}, \bibinfo {author} {\bibfnamefont {P.~V.}\ \bibnamefont
  {Klimov}}, \bibinfo {author} {\bibfnamefont {C.~F.}\ \bibnamefont {de~las
  Casas}}, \bibinfo {author} {\bibfnamefont {K.}~\bibnamefont {Sz\'asz}},
  \bibinfo {author} {\bibfnamefont {V.}~\bibnamefont {Iv\'ady}}, \bibinfo
  {author} {\bibfnamefont {V.}~\bibnamefont {Jokubavicius}}, \bibinfo {author}
  {\bibfnamefont {J.}~\bibnamefont {Ul~Hassan}}, \bibinfo {author}
  {\bibfnamefont {M.}~\bibnamefont {Syv\"aj\"arvi}}, \bibinfo {author}
  {\bibfnamefont {W.~F.}\ \bibnamefont {Koehl}}, \bibinfo {author}
  {\bibfnamefont {T.}~\bibnamefont {Ohshima}}, \bibinfo {author} {\bibfnamefont
  {N.~T.}\ \bibnamefont {Son}}, \bibinfo {author} {\bibfnamefont
  {E.}~\bibnamefont {Janz\'en}}, \bibinfo {author} {\bibfnamefont
  {A.}~\bibnamefont {Gali}}, \ and\ \bibinfo {author} {\bibfnamefont {D.~D.}\
  \bibnamefont {Awschalom}},\ }\href {\doibase 10.1103/PhysRevX.7.021046}
  {\bibfield  {journal} {\bibinfo  {journal} {Phys. Rev. X}\ }\textbf {\bibinfo
  {volume} {7}},\ \bibinfo {pages} {021046} (\bibinfo {year}
  {2017})}\BibitemShut {NoStop}%
\bibitem [{\citenamefont {Bourassa}\ \emph {et~al.}(2020)\citenamefont
  {Bourassa}, \citenamefont {Anderson}, \citenamefont {Miao}, \citenamefont
  {Onizhuk}, \citenamefont {Ma}, \citenamefont {Crook}, \citenamefont {Abe},
  \citenamefont {Ul-Hassan}, \citenamefont {Ohshima}, \citenamefont {Son},
  \citenamefont {Galli},\ and\ \citenamefont {Awschalom}}]{BourassaNatMat2020}%
  \BibitemOpen
  \bibfield  {author} {\bibinfo {author} {\bibfnamefont {A.}~\bibnamefont
  {Bourassa}}, \bibinfo {author} {\bibfnamefont {C.~P.}\ \bibnamefont
  {Anderson}}, \bibinfo {author} {\bibfnamefont {K.~C.}\ \bibnamefont {Miao}},
  \bibinfo {author} {\bibfnamefont {M.}~\bibnamefont {Onizhuk}}, \bibinfo
  {author} {\bibfnamefont {H.}~\bibnamefont {Ma}}, \bibinfo {author}
  {\bibfnamefont {A.~L.}\ \bibnamefont {Crook}}, \bibinfo {author}
  {\bibfnamefont {H.}~\bibnamefont {Abe}}, \bibinfo {author} {\bibfnamefont
  {J.}~\bibnamefont {Ul-Hassan}}, \bibinfo {author} {\bibfnamefont
  {T.}~\bibnamefont {Ohshima}}, \bibinfo {author} {\bibfnamefont {N.~T.}\
  \bibnamefont {Son}}, \bibinfo {author} {\bibfnamefont {G.}~\bibnamefont
  {Galli}}, \ and\ \bibinfo {author} {\bibfnamefont {D.~D.}\ \bibnamefont
  {Awschalom}},\ }\href {\doibase 10.1038/s41563-020-00802-6} {\bibfield
  {journal} {\bibinfo  {journal} {Nature Materials}\ }\textbf {\bibinfo
  {volume} {19}},\ \bibinfo {pages} {1319} (\bibinfo {year}
  {2020})}\BibitemShut {NoStop}%
\bibitem [{\citenamefont {Falk}\ \emph {et~al.}(2014)\citenamefont {Falk},
  \citenamefont {Klimov}, \citenamefont {Buckley}, \citenamefont {Iv\'ady},
  \citenamefont {Abrikosov}, \citenamefont {Calusine}, \citenamefont {Koehl},
  \citenamefont {Gali},\ and\ \citenamefont {Awschalom}}]{Falk2014}%
  \BibitemOpen
  \bibfield  {author} {\bibinfo {author} {\bibfnamefont {A.~L.}\ \bibnamefont
  {Falk}}, \bibinfo {author} {\bibfnamefont {P.~V.}\ \bibnamefont {Klimov}},
  \bibinfo {author} {\bibfnamefont {B.~B.}\ \bibnamefont {Buckley}}, \bibinfo
  {author} {\bibfnamefont {V.}~\bibnamefont {Iv\'ady}}, \bibinfo {author}
  {\bibfnamefont {I.~A.}\ \bibnamefont {Abrikosov}}, \bibinfo {author}
  {\bibfnamefont {G.}~\bibnamefont {Calusine}}, \bibinfo {author}
  {\bibfnamefont {W.~F.}\ \bibnamefont {Koehl}}, \bibinfo {author}
  {\bibfnamefont {A.}~\bibnamefont {Gali}}, \ and\ \bibinfo {author}
  {\bibfnamefont {D.~D.}\ \bibnamefont {Awschalom}},\ }\href {\doibase
  10.1103/PhysRevLett.112.187601} {\bibfield  {journal} {\bibinfo  {journal}
  {Physical Review Letters}\ }\textbf {\bibinfo {volume} {112}},\ \bibinfo
  {pages} {187601} (\bibinfo {year} {2014})}\BibitemShut {NoStop}%
\bibitem [{\citenamefont {Wolfowicz}\ \emph {et~al.}(2021)\citenamefont
  {Wolfowicz}, \citenamefont {Heremans}, \citenamefont {Anderson},
  \citenamefont {Kanai}, \citenamefont {Seo}, \citenamefont {Gali},
  \citenamefont {Galli},\ and\ \citenamefont {Awschalom}}]{Wolfowicz2021}%
  \BibitemOpen
  \bibfield  {author} {\bibinfo {author} {\bibfnamefont {G.}~\bibnamefont
  {Wolfowicz}}, \bibinfo {author} {\bibfnamefont {F.~J.}\ \bibnamefont
  {Heremans}}, \bibinfo {author} {\bibfnamefont {C.~P.}\ \bibnamefont
  {Anderson}}, \bibinfo {author} {\bibfnamefont {S.}~\bibnamefont {Kanai}},
  \bibinfo {author} {\bibfnamefont {H.}~\bibnamefont {Seo}}, \bibinfo {author}
  {\bibfnamefont {A.}~\bibnamefont {Gali}}, \bibinfo {author} {\bibfnamefont
  {G.}~\bibnamefont {Galli}}, \ and\ \bibinfo {author} {\bibfnamefont {D.~D.}\
  \bibnamefont {Awschalom}},\ }\href {\doibase 10.1038/s41578-021-00306-y}
  {\bibfield  {journal} {\bibinfo  {journal} {Nature Reviews Materials}\
  }\textbf {\bibinfo {volume} {6}},\ \bibinfo {pages} {906} (\bibinfo {year}
  {2021})},\ \bibinfo {note} {bandiera\_abtest: a Cg\_type: Nature Research
  Journals Number: 10 Primary\_atype: Reviews Publisher: Nature Publishing
  Group Subject\_term: Electronic devices;Quantum information;Qubits
  Subject\_term\_id: electronic-devices;quantum-information;qubits}\BibitemShut
  {NoStop}%
\bibitem [{\citenamefont {Gali}\ \emph {et~al.}(2006)\citenamefont {Gali},
  \citenamefont {Bockstedte}, \citenamefont {Son}, \citenamefont {Umeda},
  \citenamefont {Isoya},\ and\ \citenamefont {Janz{\'{e}}n}}]{GaliMSF2006}%
  \BibitemOpen
  \bibfield  {author} {\bibinfo {author} {\bibfnamefont {A.}~\bibnamefont
  {Gali}}, \bibinfo {author} {\bibfnamefont {M.}~\bibnamefont {Bockstedte}},
  \bibinfo {author} {\bibfnamefont {N.~T.}\ \bibnamefont {Son}}, \bibinfo
  {author} {\bibfnamefont {T.}~\bibnamefont {Umeda}}, \bibinfo {author}
  {\bibfnamefont {J.}~\bibnamefont {Isoya}}, \ and\ \bibinfo {author}
  {\bibfnamefont {E.}~\bibnamefont {Janz{\'{e}}n}},\ }in\ \href {\doibase
  10.4028/www.scientific.net/MSF.527-529.523} {\emph {\bibinfo {booktitle}
  {Silicon Carbide and Related Materials 2005}}},\ \bibinfo {series} {Materials
  Science Forum}, Vol.\ \bibinfo {volume} {527}\ (\bibinfo  {publisher} {Trans
  Tech Publications Ltd},\ \bibinfo {year} {2006})\ pp.\ \bibinfo {pages}
  {523--526}\BibitemShut {NoStop}%
\bibitem [{\citenamefont {Gali}(2012)}]{GaliJMR2012}%
  \BibitemOpen
  \bibfield  {author} {\bibinfo {author} {\bibfnamefont {A.}~\bibnamefont
  {Gali}},\ }\href {\doibase 10.1557/jmr.2011.431} {\bibfield  {journal}
  {\bibinfo  {journal} {Journal of Materials Research}\ }\textbf {\bibinfo
  {volume} {27}},\ \bibinfo {pages} {897â€“909} (\bibinfo {year}
  {2012})}\BibitemShut {NoStop}%
\bibitem [{\citenamefont {Gordon}\ \emph {et~al.}(2015)\citenamefont {Gordon},
  \citenamefont {Janotti},\ and\ \citenamefont {Van~de Walle}}]{GordonPRB2015}%
  \BibitemOpen
  \bibfield  {author} {\bibinfo {author} {\bibfnamefont {L.}~\bibnamefont
  {Gordon}}, \bibinfo {author} {\bibfnamefont {A.}~\bibnamefont {Janotti}}, \
  and\ \bibinfo {author} {\bibfnamefont {C.~G.}\ \bibnamefont {Van~de Walle}},\
  }\href {\doibase 10.1103/PhysRevB.92.045208} {\bibfield  {journal} {\bibinfo
  {journal} {Phys. Rev. B}\ }\textbf {\bibinfo {volume} {92}},\ \bibinfo
  {pages} {045208} (\bibinfo {year} {2015})}\BibitemShut {NoStop}%
\bibitem [{\citenamefont {Magnusson}\ \emph {et~al.}(2018)\citenamefont
  {Magnusson}, \citenamefont {Son}, \citenamefont {Cs\'or\'e}, \citenamefont
  {G\"allstr\"om}, \citenamefont {Ohshima}, \citenamefont {Gali},\ and\
  \citenamefont {Ivanov}}]{MagnussonPRB2018}%
  \BibitemOpen
  \bibfield  {author} {\bibinfo {author} {\bibfnamefont {B.}~\bibnamefont
  {Magnusson}}, \bibinfo {author} {\bibfnamefont {N.~T.}\ \bibnamefont {Son}},
  \bibinfo {author} {\bibfnamefont {A.}~\bibnamefont {Cs\'or\'e}}, \bibinfo
  {author} {\bibfnamefont {A.}~\bibnamefont {G\"allstr\"om}}, \bibinfo {author}
  {\bibfnamefont {T.}~\bibnamefont {Ohshima}}, \bibinfo {author} {\bibfnamefont
  {A.}~\bibnamefont {Gali}}, \ and\ \bibinfo {author} {\bibfnamefont {I.~G.}\
  \bibnamefont {Ivanov}},\ }\href {\doibase 10.1103/PhysRevB.98.195202}
  {\bibfield  {journal} {\bibinfo  {journal} {Phys. Rev. B}\ }\textbf {\bibinfo
  {volume} {98}},\ \bibinfo {pages} {195202} (\bibinfo {year}
  {2018})}\BibitemShut {NoStop}%
\bibitem [{\citenamefont {Anderson}\ \emph {et~al.}(2019)\citenamefont
  {Anderson}, \citenamefont {Bourassa}, \citenamefont {Miao}, \citenamefont
  {Wolfowicz}, \citenamefont {Mintun}, \citenamefont {Crook}, \citenamefont
  {Abe}, \citenamefont {Ul~Hassan}, \citenamefont {Son}, \citenamefont
  {Ohshima},\ and\ \citenamefont {Awschalom}}]{Anderson2019}%
  \BibitemOpen
  \bibfield  {author} {\bibinfo {author} {\bibfnamefont {C.~P.}\ \bibnamefont
  {Anderson}}, \bibinfo {author} {\bibfnamefont {A.}~\bibnamefont {Bourassa}},
  \bibinfo {author} {\bibfnamefont {K.~C.}\ \bibnamefont {Miao}}, \bibinfo
  {author} {\bibfnamefont {G.}~\bibnamefont {Wolfowicz}}, \bibinfo {author}
  {\bibfnamefont {P.~J.}\ \bibnamefont {Mintun}}, \bibinfo {author}
  {\bibfnamefont {A.~L.}\ \bibnamefont {Crook}}, \bibinfo {author}
  {\bibfnamefont {H.}~\bibnamefont {Abe}}, \bibinfo {author} {\bibfnamefont
  {J.}~\bibnamefont {Ul~Hassan}}, \bibinfo {author} {\bibfnamefont {N.~T.}\
  \bibnamefont {Son}}, \bibinfo {author} {\bibfnamefont {T.}~\bibnamefont
  {Ohshima}}, \ and\ \bibinfo {author} {\bibfnamefont {D.~D.}\ \bibnamefont
  {Awschalom}},\ }\href {\doibase 10.1126/science.aax9406} {\bibfield
  {journal} {\bibinfo  {journal} {Science}\ }\textbf {\bibinfo {volume}
  {366}},\ \bibinfo {pages} {1225} (\bibinfo {year} {2019})},\ \bibinfo {note}
  {publisher: American Association for the Advancement of Science}\BibitemShut
  {NoStop}%
\bibitem [{\citenamefont {Zwier}\ \emph {et~al.}(2015)\citenamefont {Zwier},
  \citenamefont {O'Shea}, \citenamefont {Onur},\ and\ \citenamefont {van~der
  Wal}}]{ZwierSciRep2015}%
  \BibitemOpen
  \bibfield  {author} {\bibinfo {author} {\bibfnamefont {O.}~\bibnamefont
  {Zwier}}, \bibinfo {author} {\bibfnamefont {D.}~\bibnamefont {O'Shea}},
  \bibinfo {author} {\bibfnamefont {A.~R.}\ \bibnamefont {Onur}}, \ and\
  \bibinfo {author} {\bibfnamefont {C.~H.}\ \bibnamefont {van~der Wal}},\
  }\href {\doibase 10.1038/srep10931} {\bibfield  {journal} {\bibinfo
  {journal} {Sci. Rep.}\ }\textbf {\bibinfo {volume} {5}} (\bibinfo {year}
  {2015}),\ 10.1038/srep10931}\BibitemShut {NoStop}%
\bibitem [{\citenamefont {Wolfowicz}\ \emph {et~al.}(2017)\citenamefont
  {Wolfowicz}, \citenamefont {Anderson}, \citenamefont {Yeats}, \citenamefont
  {Whiteley}, \citenamefont {Niklas}, \citenamefont {Poluektov}, \citenamefont
  {Heremans},\ and\ \citenamefont {Awschalom}}]{WolfowiczNatComm2017}%
  \BibitemOpen
  \bibfield  {author} {\bibinfo {author} {\bibfnamefont {G.}~\bibnamefont
  {Wolfowicz}}, \bibinfo {author} {\bibfnamefont {C.~P.}\ \bibnamefont
  {Anderson}}, \bibinfo {author} {\bibfnamefont {A.~L.}\ \bibnamefont {Yeats}},
  \bibinfo {author} {\bibfnamefont {S.~J.}\ \bibnamefont {Whiteley}}, \bibinfo
  {author} {\bibfnamefont {J.}~\bibnamefont {Niklas}}, \bibinfo {author}
  {\bibfnamefont {O.~G.}\ \bibnamefont {Poluektov}}, \bibinfo {author}
  {\bibfnamefont {F.~J.}\ \bibnamefont {Heremans}}, \ and\ \bibinfo {author}
  {\bibfnamefont {D.~D.}\ \bibnamefont {Awschalom}},\ }\href {\doibase
  10.1038/s41467-017-01993-4} {\bibfield  {journal} {\bibinfo  {journal} {Nat.
  Comm.}\ }\textbf {\bibinfo {volume} {8}} (\bibinfo {year} {2017}),\
  10.1038/s41467-017-01993-4}\BibitemShut {NoStop}%
\bibitem [{\citenamefont {Golter}\ and\ \citenamefont
  {Lai}(2017)}]{GolterSciRep2017}%
  \BibitemOpen
  \bibfield  {author} {\bibinfo {author} {\bibfnamefont {D.~A.}\ \bibnamefont
  {Golter}}\ and\ \bibinfo {author} {\bibfnamefont {C.~W.}\ \bibnamefont
  {Lai}},\ }\href {\doibase 10.1038/s41598-017-13813-2} {\bibfield  {journal}
  {\bibinfo  {journal} {Sci. Rep.}\ }\textbf {\bibinfo {volume} {7}} (\bibinfo
  {year} {2017}),\ 10.1038/s41598-017-13813-2}\BibitemShut {NoStop}%
\bibitem [{\citenamefont {Siyushev}\ \emph {et~al.}(2013)\citenamefont
  {Siyushev}, \citenamefont {Pinto}, \citenamefont {V\"or\"os}, \citenamefont
  {Gali}, \citenamefont {Jelezko},\ and\ \citenamefont
  {Wrachtrup}}]{SiyushevPRL2013}%
  \BibitemOpen
  \bibfield  {author} {\bibinfo {author} {\bibfnamefont {P.}~\bibnamefont
  {Siyushev}}, \bibinfo {author} {\bibfnamefont {H.}~\bibnamefont {Pinto}},
  \bibinfo {author} {\bibfnamefont {M.}~\bibnamefont {V\"or\"os}}, \bibinfo
  {author} {\bibfnamefont {A.}~\bibnamefont {Gali}}, \bibinfo {author}
  {\bibfnamefont {F.}~\bibnamefont {Jelezko}}, \ and\ \bibinfo {author}
  {\bibfnamefont {J.}~\bibnamefont {Wrachtrup}},\ }\href {\doibase
  10.1103/PhysRevLett.110.167402} {\bibfield  {journal} {\bibinfo  {journal}
  {Phys. Rev. Lett.}\ }\textbf {\bibinfo {volume} {110}},\ \bibinfo {pages}
  {167402} (\bibinfo {year} {2013})}\BibitemShut {NoStop}%
\bibitem [{\citenamefont {Huang}\ \emph {et~al.}(1950)\citenamefont {Huang},
  \citenamefont {Rhys},\ and\ \citenamefont {Mott}}]{HuangProcRoySoC1950}%
  \BibitemOpen
  \bibfield  {author} {\bibinfo {author} {\bibfnamefont {K.}~\bibnamefont
  {Huang}}, \bibinfo {author} {\bibfnamefont {A.}~\bibnamefont {Rhys}}, \ and\
  \bibinfo {author} {\bibfnamefont {N.~F.}\ \bibnamefont {Mott}},\ }\href
  {\doibase 10.1098/rspa.1950.0184} {\bibfield  {journal} {\bibinfo  {journal}
  {Proceedings of the Royal Society of London. Series A. Mathematical and
  Physical Sciences}\ }\textbf {\bibinfo {volume} {204}},\ \bibinfo {pages}
  {406} (\bibinfo {year} {1950})},\ \Eprint
  {http://arxiv.org/abs/https://royalsocietypublishing.org/doi/pdf/10.1098/rspa.1950.0184}
  {https://royalsocietypublishing.org/doi/pdf/10.1098/rspa.1950.0184}
  \BibitemShut {NoStop}%
\bibitem [{\citenamefont {Alkauskas}\ \emph {et~al.}(2014)\citenamefont
  {Alkauskas}, \citenamefont {Buckley}, \citenamefont {Awschalom},\ and\
  \citenamefont {de~Walle}}]{AlkauskasNJP2014}%
  \BibitemOpen
  \bibfield  {author} {\bibinfo {author} {\bibfnamefont {A.}~\bibnamefont
  {Alkauskas}}, \bibinfo {author} {\bibfnamefont {B.~B.}\ \bibnamefont
  {Buckley}}, \bibinfo {author} {\bibfnamefont {D.~D.}\ \bibnamefont
  {Awschalom}}, \ and\ \bibinfo {author} {\bibfnamefont {C.~G.~V.}\
  \bibnamefont {de~Walle}},\ }\href
  {http://stacks.iop.org/1367-2630/16/i=7/a=073026} {\bibfield  {journal}
  {\bibinfo  {journal} {New Journal of Physics}\ }\textbf {\bibinfo {volume}
  {16}},\ \bibinfo {pages} {073026} (\bibinfo {year} {2014})}\BibitemShut
  {NoStop}%
\bibitem [{\citenamefont {Gali}\ \emph {et~al.}(2016)\citenamefont {Gali},
  \citenamefont {T.}, \citenamefont {V\"or\"os}, \citenamefont {Thiering},
  \citenamefont {Cannuccia},\ and\ \citenamefont {Marini}}]{GaliNatComm2016}%
  \BibitemOpen
  \bibfield  {author} {\bibinfo {author} {\bibfnamefont {A.}~\bibnamefont
  {Gali}}, \bibinfo {author} {\bibfnamefont {D.}~\bibnamefont {T.}}, \bibinfo
  {author} {\bibfnamefont {M.}~\bibnamefont {V\"or\"os}}, \bibinfo {author}
  {\bibfnamefont {G.}~\bibnamefont {Thiering}}, \bibinfo {author}
  {\bibfnamefont {E.}~\bibnamefont {Cannuccia}}, \ and\ \bibinfo {author}
  {\bibfnamefont {A.}~\bibnamefont {Marini}},\ }\href {\doibase
  10.1038/ncomms11327} {\bibfield  {journal} {\bibinfo  {journal} {Nat. Comm.}\
  }\textbf {\bibinfo {volume} {7}} (\bibinfo {year} {2016}),\
  10.1038/ncomms11327}\BibitemShut {NoStop}%
\bibitem [{\citenamefont {{Debye}}(1913)}]{DebyeAdP1913}%
  \BibitemOpen
  \bibfield  {author} {\bibinfo {author} {\bibfnamefont {P.}~\bibnamefont
  {{Debye}}},\ }\href {\doibase 10.1002/andp.19133480105} {\bibfield  {journal}
  {\bibinfo  {journal} {Annalen der Physik}\ }\textbf {\bibinfo {volume}
  {348}},\ \bibinfo {pages} {49} (\bibinfo {year} {1913})}\BibitemShut
  {NoStop}%
\bibitem [{\citenamefont {{Waller}}(1923)}]{WallerZfP1923}%
  \BibitemOpen
  \bibfield  {author} {\bibinfo {author} {\bibfnamefont {I.}~\bibnamefont
  {{Waller}}},\ }\href {\doibase 10.1007/BF01328696} {\bibfield  {journal}
  {\bibinfo  {journal} {Zeitschrift fur Physik}\ }\textbf {\bibinfo {volume}
  {17}},\ \bibinfo {pages} {398} (\bibinfo {year} {1923})}\BibitemShut
  {NoStop}%
\bibitem [{\citenamefont {Cs\'or\'e}\ \emph {et~al.}(2021)\citenamefont
  {Cs\'or\'e}, \citenamefont {Son},\ and\ \citenamefont {Gali}}]{CsorePRB2021}%
  \BibitemOpen
  \bibfield  {author} {\bibinfo {author} {\bibfnamefont {A.}~\bibnamefont
  {Cs\'or\'e}}, \bibinfo {author} {\bibfnamefont {N.~T.}\ \bibnamefont {Son}},
  \ and\ \bibinfo {author} {\bibfnamefont {A.}~\bibnamefont {Gali}},\ }\href
  {\doibase 10.1103/PhysRevB.104.035207} {\bibfield  {journal} {\bibinfo
  {journal} {Phys. Rev. B}\ }\textbf {\bibinfo {volume} {104}},\ \bibinfo
  {pages} {035207} (\bibinfo {year} {2021})}\BibitemShut {NoStop}%
\bibitem [{\citenamefont {Freysoldt}\ \emph {et~al.}(2009)\citenamefont
  {Freysoldt}, \citenamefont {Neugebauer},\ and\ \citenamefont {Van~de
  Walle}}]{FreysoldtPRL2009}%
  \BibitemOpen
  \bibfield  {author} {\bibinfo {author} {\bibfnamefont {C.}~\bibnamefont
  {Freysoldt}}, \bibinfo {author} {\bibfnamefont {J.}~\bibnamefont
  {Neugebauer}}, \ and\ \bibinfo {author} {\bibfnamefont {C.~G.}\ \bibnamefont
  {Van~de Walle}},\ }\href {\doibase 10.1103/PhysRevLett.102.016402} {\bibfield
   {journal} {\bibinfo  {journal} {Phys. Rev. Lett.}\ }\textbf {\bibinfo
  {volume} {102}},\ \bibinfo {pages} {016402} (\bibinfo {year}
  {2009})}\BibitemShut {NoStop}%
\bibitem [{\citenamefont {Cannuccia}\ and\ \citenamefont
  {Gali}(2020)}]{CannucciaPRMat2020}%
  \BibitemOpen
  \bibfield  {author} {\bibinfo {author} {\bibfnamefont {E.}~\bibnamefont
  {Cannuccia}}\ and\ \bibinfo {author} {\bibfnamefont {A.}~\bibnamefont
  {Gali}},\ }\href {\doibase 10.1103/PhysRevMaterials.4.014601} {\bibfield
  {journal} {\bibinfo  {journal} {Phys. Rev. Materials}\ }\textbf {\bibinfo
  {volume} {4}},\ \bibinfo {pages} {014601} (\bibinfo {year}
  {2020})}\BibitemShut {NoStop}%
\bibitem [{\citenamefont {Wickramaratne}\ \emph {et~al.}(2018)\citenamefont
  {Wickramaratne}, \citenamefont {Dreyer}, \citenamefont {Monserrat},
  \citenamefont {Shen}, \citenamefont {Lyons}, \citenamefont {Alkauskas},\ and\
  \citenamefont {Van~de Walle}}]{WickamaranteAPL2018}%
  \BibitemOpen
  \bibfield  {author} {\bibinfo {author} {\bibfnamefont {D.}~\bibnamefont
  {Wickramaratne}}, \bibinfo {author} {\bibfnamefont {C.~E.}\ \bibnamefont
  {Dreyer}}, \bibinfo {author} {\bibfnamefont {B.}~\bibnamefont {Monserrat}},
  \bibinfo {author} {\bibfnamefont {J.-X.}\ \bibnamefont {Shen}}, \bibinfo
  {author} {\bibfnamefont {J.~L.}\ \bibnamefont {Lyons}}, \bibinfo {author}
  {\bibfnamefont {A.}~\bibnamefont {Alkauskas}}, \ and\ \bibinfo {author}
  {\bibfnamefont {C.~G.}\ \bibnamefont {Van~de Walle}},\ }\href {\doibase
  10.1063/1.5047808} {\bibfield  {journal} {\bibinfo  {journal} {Applied
  Physics Letters}\ }\textbf {\bibinfo {volume} {113}},\ \bibinfo {pages}
  {192106} (\bibinfo {year} {2018})},\ \Eprint
  {http://arxiv.org/abs/https://doi.org/10.1063/1.5047808}
  {https://doi.org/10.1063/1.5047808} \BibitemShut {NoStop}%
\bibitem [{\citenamefont {Gali}\ \emph {et~al.}(2009)\citenamefont {Gali},
  \citenamefont {Janz\'en}, \citenamefont {De\'ak}, \citenamefont {Kresse},\
  and\ \citenamefont {Kaxiras}}]{GaliPRL2009}%
  \BibitemOpen
  \bibfield  {author} {\bibinfo {author} {\bibfnamefont {A.}~\bibnamefont
  {Gali}}, \bibinfo {author} {\bibfnamefont {E.}~\bibnamefont {Janz\'en}},
  \bibinfo {author} {\bibfnamefont {P.}~\bibnamefont {De\'ak}}, \bibinfo
  {author} {\bibfnamefont {G.}~\bibnamefont {Kresse}}, \ and\ \bibinfo {author}
  {\bibfnamefont {E.}~\bibnamefont {Kaxiras}},\ }\href {\doibase
  10.1103/PhysRevLett.103.186404} {\bibfield  {journal} {\bibinfo  {journal}
  {Phys. Rev. Lett.}\ }\textbf {\bibinfo {volume} {103}},\ \bibinfo {pages}
  {186404} (\bibinfo {year} {2009})}\BibitemShut {NoStop}%
\bibitem [{\citenamefont {Bersuker}(2006)}]{Bersuker2006}%
  \BibitemOpen
  \bibfield  {author} {\bibinfo {author} {\bibfnamefont {I.~B.}\ \bibnamefont
  {Bersuker}},\ }\href@noop {} {\emph {\bibinfo {title} {The Jahn-Teller
  effect}}}\ (\bibinfo  {publisher} {Cambridge University Press},\ \bibinfo
  {year} {2006})\BibitemShut {NoStop}%
\bibitem [{\citenamefont {Heyd}\ \emph {et~al.}(2003)\citenamefont {Heyd},
  \citenamefont {Scuseria},\ and\ \citenamefont {Ernzerhof}}]{HeydJCP2003}%
  \BibitemOpen
  \bibfield  {author} {\bibinfo {author} {\bibfnamefont {J.}~\bibnamefont
  {Heyd}}, \bibinfo {author} {\bibfnamefont {G.~E.}\ \bibnamefont {Scuseria}},
  \ and\ \bibinfo {author} {\bibfnamefont {M.}~\bibnamefont {Ernzerhof}},\
  }\href {\doibase http://dx.doi.org/10.1063/1.1564060} {\bibfield  {journal}
  {\bibinfo  {journal} {The Journal of Chemical Physics}\ }\textbf {\bibinfo
  {volume} {118}},\ \bibinfo {pages} {8207} (\bibinfo {year}
  {2003})}\BibitemShut {NoStop}%
\bibitem [{\citenamefont {Bl\"ochl}(1994)}]{BlochlPRB1994}%
  \BibitemOpen
  \bibfield  {author} {\bibinfo {author} {\bibfnamefont {P.~E.}\ \bibnamefont
  {Bl\"ochl}},\ }\href {\doibase 10.1103/PhysRevB.50.17953} {\bibfield
  {journal} {\bibinfo  {journal} {Phys. Rev. B}\ }\textbf {\bibinfo {volume}
  {50}},\ \bibinfo {pages} {17953} (\bibinfo {year} {1994})}\BibitemShut
  {NoStop}%
\bibitem [{\citenamefont {Kresse}\ and\ \citenamefont
  {Furthm\"uller}(1996)}]{KressePRB1996}%
  \BibitemOpen
  \bibfield  {author} {\bibinfo {author} {\bibfnamefont {G.}~\bibnamefont
  {Kresse}}\ and\ \bibinfo {author} {\bibfnamefont {J.}~\bibnamefont
  {Furthm\"uller}},\ }\href {\doibase 10.1103/PhysRevB.54.11169} {\bibfield
  {journal} {\bibinfo  {journal} {Phys. Rev. B}\ }\textbf {\bibinfo {volume}
  {54}},\ \bibinfo {pages} {11169} (\bibinfo {year} {1996})}\BibitemShut
  {NoStop}%
\bibitem [{\citenamefont {Perdew}\ \emph {et~al.}(1996)\citenamefont {Perdew},
  \citenamefont {Burke},\ and\ \citenamefont {Ernzerhof}}]{PerdewPRL1996}%
  \BibitemOpen
  \bibfield  {author} {\bibinfo {author} {\bibfnamefont {J.~P.}\ \bibnamefont
  {Perdew}}, \bibinfo {author} {\bibfnamefont {K.}~\bibnamefont {Burke}}, \
  and\ \bibinfo {author} {\bibfnamefont {M.}~\bibnamefont {Ernzerhof}},\ }\href
  {\doibase 10.1103/PhysRevLett.77.3865} {\bibfield  {journal} {\bibinfo
  {journal} {Phys. Rev. Lett.}\ }\textbf {\bibinfo {volume} {77}},\ \bibinfo
  {pages} {3865} (\bibinfo {year} {1996})}\BibitemShut {NoStop}%
\bibitem [{\citenamefont {Bell}\ \emph {et~al.}(1970)\citenamefont {Bell},
  \citenamefont {Dean},\ and\ \citenamefont {Hibbins-Butler}}]{Bell1970JOP}%
  \BibitemOpen
  \bibfield  {author} {\bibinfo {author} {\bibfnamefont {R.~J.}\ \bibnamefont
  {Bell}}, \bibinfo {author} {\bibfnamefont {P.}~\bibnamefont {Dean}}, \ and\
  \bibinfo {author} {\bibfnamefont {D.~C.}\ \bibnamefont {Hibbins-Butler}},\
  }\href {\doibase 10.1088/0022-3719/3/10/013} {\bibfield  {journal} {\bibinfo
  {journal} {Journal of Physics C: Solid State Physics}\ }\textbf {\bibinfo
  {volume} {3}},\ \bibinfo {pages} {2111} (\bibinfo {year} {1970})}\BibitemShut
  {NoStop}%
\bibitem [{\citenamefont {Falk}\ \emph {et~al.}(2013)\citenamefont {Falk},
  \citenamefont {Buckley}, \citenamefont {Calusine}, \citenamefont {Koehl},
  \citenamefont {Dobrovitski}, \citenamefont {Politi}, \citenamefont {Zorman},
  \citenamefont {Feng},\ and\ \citenamefont {Awschalom}}]{Falk2013NatCom}%
  \BibitemOpen
  \bibfield  {author} {\bibinfo {author} {\bibfnamefont {A.~L.}\ \bibnamefont
  {Falk}}, \bibinfo {author} {\bibfnamefont {B.~B.}\ \bibnamefont {Buckley}},
  \bibinfo {author} {\bibfnamefont {G.}~\bibnamefont {Calusine}}, \bibinfo
  {author} {\bibfnamefont {W.~F.}\ \bibnamefont {Koehl}}, \bibinfo {author}
  {\bibfnamefont {V.~V.}\ \bibnamefont {Dobrovitski}}, \bibinfo {author}
  {\bibfnamefont {A.}~\bibnamefont {Politi}}, \bibinfo {author} {\bibfnamefont
  {C.~A.}\ \bibnamefont {Zorman}}, \bibinfo {author} {\bibfnamefont {P.~X.-L.}\
  \bibnamefont {Feng}}, \ and\ \bibinfo {author} {\bibfnamefont {D.~D.}\
  \bibnamefont {Awschalom}},\ }\href {\doibase 10.1038/ncomms2854} {\bibfield
  {journal} {\bibinfo  {journal} {Nat Commun}\ }\textbf {\bibinfo {volume}
  {4}},\ \bibinfo {pages} {1819} (\bibinfo {year} {2013})}\BibitemShut
  {NoStop}%
\bibitem [{\citenamefont {Hashemi}\ \emph {et~al.}(2021)\citenamefont
  {Hashemi}, \citenamefont {Linder\"alv}, \citenamefont {Krasheninnikov},
  \citenamefont {Ala-Nissila}, \citenamefont {Erhart},\ and\ \citenamefont
  {Komsa}}]{HashemiPRB2021}%
  \BibitemOpen
  \bibfield  {author} {\bibinfo {author} {\bibfnamefont {A.}~\bibnamefont
  {Hashemi}}, \bibinfo {author} {\bibfnamefont {C.}~\bibnamefont
  {Linder\"alv}}, \bibinfo {author} {\bibfnamefont {A.~V.}\ \bibnamefont
  {Krasheninnikov}}, \bibinfo {author} {\bibfnamefont {T.}~\bibnamefont
  {Ala-Nissila}}, \bibinfo {author} {\bibfnamefont {P.}~\bibnamefont {Erhart}},
  \ and\ \bibinfo {author} {\bibfnamefont {H.-P.}\ \bibnamefont {Komsa}},\
  }\href {\doibase 10.1103/PhysRevB.103.125203} {\bibfield  {journal} {\bibinfo
   {journal} {Phys. Rev. B}\ }\textbf {\bibinfo {volume} {103}},\ \bibinfo
  {pages} {125203} (\bibinfo {year} {2021})}\BibitemShut {NoStop}%
\bibitem [{\citenamefont {Cs\'or\'e}\ \emph {et~al.}(2022)\citenamefont
  {Cs\'or\'e}, \citenamefont {Mukesh}, \citenamefont {K\'arolyh\'azy},
  \citenamefont {Beke},\ and\ \citenamefont {Gali}}]{Csore2022}%
  \BibitemOpen
  \bibfield  {author} {\bibinfo {author} {\bibfnamefont {A.}~\bibnamefont
  {Cs\'or\'e}}, \bibinfo {author} {\bibfnamefont {N.}~\bibnamefont {Mukesh}},
  \bibinfo {author} {\bibfnamefont {G.}~\bibnamefont {K\'arolyh\'azy}},
  \bibinfo {author} {\bibfnamefont {D.}~\bibnamefont {Beke}}, \ and\ \bibinfo
  {author} {\bibfnamefont {A.}~\bibnamefont {Gali}},\ }\href {\doibase
  10.1063/5.0080514} {\bibfield  {journal} {\bibinfo  {journal} {Journal of
  Applied Physics}\ }\textbf {\bibinfo {volume} {131}},\ \bibinfo {pages}
  {071102} (\bibinfo {year} {2022})},\ \bibinfo {note} {publisher: American
  Institute of Physics}\BibitemShut {NoStop}%
\bibitem [{\citenamefont {Lax}(1960)}]{LaxPR1960}%
  \BibitemOpen
  \bibfield  {author} {\bibinfo {author} {\bibfnamefont {M.}~\bibnamefont
  {Lax}},\ }\href {\doibase 10.1103/PhysRev.119.1502} {\bibfield  {journal}
  {\bibinfo  {journal} {Phys. Rev.}\ }\textbf {\bibinfo {volume} {119}},\
  \bibinfo {pages} {1502} (\bibinfo {year} {1960})}\BibitemShut {NoStop}%
\end{thebibliography}

%

\end{document}